\documentclass[a4paper,fleqn,usenatbib]{mnras}

\usepackage{newtxtext,newtxmath}
\usepackage[T1]{fontenc}
\usepackage{ae,aecompl}
\usepackage{listings}

\usepackage{graphicx}	% Including figure files
\usepackage{color}      % use if color is used in text
\usepackage{float}
\usepackage{caption}
\usepackage{subcaption}
\usepackage{ae,aecompl,bm,booktabs}
\usepackage{amsmath,color,xspace,multirow,longtable,graphicx,bigints,comment}
\usepackage{tabularx}
\usepackage{pdflscape}

\graphicspath{{./Graphs/}}
\def\measurement setSFR{\mathrel{\langle \rm sSFR \rangle}}

\definecolor{purple}{RGB}{175,0,175}
\definecolor{red}{RGB}{255,0,0}
\definecolor{darkblue}{RGB}{0,0,175}
\definecolor{lime}{RGB}{0,255,0}

\title[Cold gas halos around ERQs]{Evidence of extended cold molecular gas and dust halos around $\mathbf{z\sim2.3}$ Extremely Red Quasars with ALMA}

\author[J. Scholtz, et al.]{\parbox[h]{\textwidth}{ 
J.\ Scholtz$,^{\! 1,2}$\thanks{E-mail: js2685@cam.ac.uk}
R. Maiolino$,^{1,2}$
G.~C. Jones$,^{3}$
S. Carniani$^{4}$
}
\\
\\
$^{1}$Cavendish Laboratory, University of Cambridge, 19 J. J. Thomson Ave., Cambridge CB3 0HE, UK;\\
$^{2}$Kavli Institute for Cosmology, University of Cambridge, Madingley Road, Cambridge CB3 0HA, UK\\
$^{3}$Department of Physics, University of Oxford, Denys Wilkinson Building, Keble Road, Oxford OX1 3RH, UK\\
$^{4}$Scuola Normale Superiore, Piazza dei Cavalieri 7, I-56126 Pisa, Italy\\
}

\date{XYZ}

\pubyear{2022}

\begin{document}
\label{firstpage}
\pagerange{\pageref{firstpage}--\pageref{lastpage}}
\maketitle

\begin{abstract}
Large-scale outflows are believed to be an important mechanism in the evolution of galaxies. We can determine the impact of these outflows by studying either  current galaxy outflows and their effect in the galaxy or by studying the effect of past outflows on the gas surrounding the galaxy. In this work, we examine the CO(7-6), [\ion{C}{i}]\,($^{3} \rm P_{1} \rightarrow {\rm ^3 P}_{0}$), H$_2$O 2$_{11}$--2$_{02}$ and dust continuum emission of 15 extremely red quasars (ERQs) at z$\sim$2.3 using ALMA. By investigating the radial surface brightness profiles of both the individual sources and the stacked emission, we detect extended cold gas and dust emission on scales of $\sim$14 kpc in CO(7-6), [\ion{C}{i}](2-1), and dust continuum. This is the first time that the presence of a large amount of molecular gas was detected on large, circum-galactic medium scales around quasar host galaxies using [\ion{C}{i}] extended emission. We estimate the dust and molecular gas mass of these halos to be 10$^{7.6}$ and 10$^{10.6}$ M$_\odot$, indicating significant dust and molecular gas reservoirs around these extreme quasars. By estimating the timescale at which this gas can reach these distances by molecular gas outflows (7-32 Myr), we conclude that these halos are a relic of past AGN or starburst activity, rather than an effect of the current episode of extreme quasar activity.

\end{abstract}

\begin{keywords}galaxies: haloes - galaxies: high-redshift - galaxies: evolution - ISM \end{keywords}

%%%%%%%%%%%%%%%%%%%%%%%%%%%%%%%%%%%%%%%%%%%%%%%%%%

%%%%%%%%%%%%%%%%% BODY OF PAPER %%%%%%%%%%%%%%%%%%

\section{Introduction}

During their growth via accretion, supermassive black holes become visible as active galactic nuclei \citep[AGN;][]{Rees82, LyndenBell69, Soltan82,Merloni04}. The feedback from AGN is a key ingredient in cosmological simulations, as it injects energy into the host galaxies and circumgalactic medium (CGM). This is required in order to reproduce key galaxy properties including the M-sigma relationship, colour bi-modality, enrichment of the of the IGM by metals, galaxy sizes and broader range of specific star formation rates \citep[e.g.,][]{Silk98,DiMatteo05,Alexander12,Dubois13,Dubois13b,Vogelsberger14,Hirschmann14, Crain15,Segers16,Beckmann17,Harrison17,Choi18,Scholtz18}.

The  CGM surrounding a galaxy is spatially extended material (on scales larger than the galaxy's stellar size; $>8$\,kpc at $z\sim2$; \citealt{ForsterSch18a}) that serves as a fuel reservoir for future star formation. This reservoir contains both the processed gas ejected from the galaxy via star-formation feedback \citep[e.g.][]{Ginolfi17,Spilker18,Gallerani18,Jones19} or AGN feedback \citep[e.g.][]{Bischetti19, Vayner21,Travascio20,Cicone21} as well as streams of gas from the large scale filaments \citep[e.g.][]{ArrigoniBattaia18, Umehata19}. Therefore, it can be used to study past star formation and AGN activity as well as future fuelling of galaxy growth.

The most reliable efforts to map the CGM come from integral-field-spectroscopy observations of Ly$\alpha$ emission, which traces ionised gas \citep[e.g.,][]{drak20,sand21}. These observations constrained the size of these ionised gas halos up to $\sim170$\,kpc. However, Ly$\alpha$ observations are tracing only a single phase of the gas in the CGM, omitting the cold phase.

Cold gas halos have been detected in both stacked [\ion{C}{ii}] 158 $\mu$m emission and individual star-forming main sequence galaxies \citep[SFMS; ][]{Schreiber15} at $z>4$ \citep[][]{Fujimoto19,Fujimoto20, Herrera-Camus21, Lambert22} that extend on scales larger than both the beam of the observations and UV emission tracing the young stars. These extended halos are taken as evidence of enrichment of the CGM by starbursts driven outflows \citep[see][for more details]{Ginolfi20}. However, in these observations, AGN were purposefully discarded to study the effect of starburst-driven outflows on the CGM.

Although we have a wealth of observations of cold gas halos, the picture around AGN host galaxies is still emerging. More recently, two studies have detected CO(3-2) halos around X-ray AGN at $z\sim$ 2.4 (\citealt{Cicone21} in a single AGN host galaxy, \citealt{Jones22} in stacked emission). In radio loud AGN, cold gas molecular halos have been detected in the Spiderweb galaxy, a galaxy protocluster at $z\sim$2.2, with a size of up to 70 kpc using CO(4-3), [\ion{C}{i}](1-0),  [\ion{C}{ii}] emission \citep[][]{Emonts+16, Emonts+18, DeBreuck+22}, and in a radio loud quasar at $z\sim$2.2, where the molecular gas reservoir is aligned with a radio jet on scales of 100 kpc \citep[][]{Li+21}. However, we still lack a systematic study of cold gas halos around quasar host galaxies.

Extremely red quasars (ERQs), a population of unique obscured quasars, are often luminous dust-reddened sources that are believed to be at a different evolutionary stage compared to blue quasars and moderate luminosity AGN \citep[][]{Ross15, Hamann17, Klindt+19}. %They are believed to be a transitional stage between a starburst and unobscured QSOs, making them an important stage in galaxy evolution. 
ERQs were initially discovered through a selection based on high rest-frame ultraviolet to infrared colours (i-W3$>$ 4.6 mag) from the Wide-Field Infrared Survey Explorer \citep[WISE; ][]{Wright10} and Sloan Digital Sky survey (SDSS; \citealt{Blanton+17}). This selection results in high nuclear obscuration with column densities up to 10$^{24}$ cm$^{-2}$ \citep[][]{Goulding18, Ishikawa21}.

Previous detections of cold gas halos have used [\ion{C}{ii}] emission, which is tracing multiple phases of gas, or CO(3-2), which requires multiple conversion factors to estimate molecular gas mass (e.g., $r_{J1}$ and $\alpha_{CO}$; \citealt{Bolatto13}). In this work we focus on analysing ALMA observations of three emission lines: CO(7-6), [\ion{C}{i}]\,($^{3} \rm P_{2} \rightarrow {\rm ^3 P}_{1}$) and H$_2$O 2$_{11}$--2$_{02}$, as well as the dust continuum (rest-frame $\sim 350 \mu$m). The CO(7-6) and [\ion{C}{i}](2-1) are among the most luminous emission lines, and therefore important lines to cool the ISM and CGM, each tracing a different cold gas phase. Similarly to CO(1-0), the [\ion{C}{i}] emission line is tracing primarily molecular gas from the outer layers of molecular clouds \citep[see][]{Walter11,AlaghbandZadeh16,Glover16,Papadopoulos18,Valentino18,Bisbas19} as well as the diffuse molecular gas in the interstellar medium, such as photon dominated regions \citep[PDRs;][]{Tielens85} with relatively low critical density ($n_{\rm crit}$) of  $\sim 10^{3}$ cm$^{-3}$. Hence [\ion{C}{i}] is an excellent trace of the cold molecular gas \citep[][]{Bothwell17, Jiao17}. Furthermore, the conversion of [\ion{C}{i}] integrated line luminosity to H$_2$ mass has smaller uncertainties compared to that of CO, and has a lesser dependency on the metallicity of the gas. On the other hand, the CO(7-6) is tracing warmer and denser gas ($T_{\rm ex}\sim$150 K and $n_{\rm crit}\sim 10^{5}$ cm$^{-3}$). This emission line is an excellent tracer of current or recent star formation \citep[see][]{Lu18,Zhao20} in star-forming galaxies, but the exact source of excitation in quasar host galaxies is yet unknown. %[https://iopscience.iop.org/article/10.3847/1538-4357/ab75eb ]

In \S~\ref{sec:data} we describe our targets, and the observations used in our study, while \S~\ref{sec:analyses} describes the data analyses of the ALMA observations, including spectral fitting, stacking and modelling the extracted radial surface brightness profile.
In \S~\ref{sec:results} we present our results and discuss the physical properties and origins of the discovered cold gas and dust halos. In this work, we consider any additional component large than the galaxy as a halo. 
In all of our analyses, we adopt the cosmological parameters of
$H_{\rm0} = 67.3$\,km\,s$^{-1}$, $\rm \Omega_M = 0.3$, $\rm
\Omega_\Lambda = 0.7$ \citep{Planck13} and assume a \citet{Chabrier03} initial mass
function (IMF).

\section{Data and observations}\label{sec:data}

\subsection{ALMA Observations and data reduction}
 In this project we investigate the molecular gas emission of fifteen extremely red quasars (ERQs) at $z\sim$ 2.4, observed in CO(7-6), [\ion{C}{i}]\,($^{3} \rm P_{2} \rightarrow {\rm ^3 P}_{1}$) and H$_2$O 2$_{11}$--2$_{02}$ emission and 1.2 mm dust continuum in ALMA project 2017.1.00478.S (PI: Fred Hamann) in Cycle 5. These ERQs were identified in \citet{Hamann17} with the following selection criteria: 1) QSOs with red colours (i-W3 $>$ 4.6); 2) large equivalent with of the CIV emission line (RW(CIV) $>$ 100 \AA ). The IDs, coordinates, redshifts, black hole masses and bolometric luminosities of the sample are summarised in Table \ref{Table:Sample}. 
 
The raw science data models were processed by the ALMA staff through their calibration pipeline, which resulted in the calibrated measurement sets for each observation. As two sources were observed over two separate executions, we joined these executions using CASA's \texttt{concat} task. The measurement sets included also the calibrator sources,  so we created measurement sets containing only calibrated visibilities of the target source with CASA's \texttt{split} task. 

With the measurement sets for each source, we imaged the sources using a uniform imaging pipeline. As the optical and sub-mm line emission can have a significant velocity offsets, we first imaged the cubes using the natural weighting (to maximise the sensitivity) to find the redshift of the CO(7-6), [\ion{C}{i}]\,($^{3} \rm P_{2} \rightarrow {\rm ^3 P}_{1}$) (hereafter [\ion{C}{i}](2-1)) and H$_2$O 2$_{11}$--2$_{02}$ (752.033 GHz; hereafter H$_2$O) emission lines. Indeed we find that the velocity offset between the optical and CO redshifts is up to 2000\, km\,s$^{-1}$. To identify the ``line-fre'' channels we picked channels outside of the $\pm600$\, km\,s$^{-1}$ of the centre of the detected emission lines. We then used CASA's \texttt{uvcontsub} task to subtract the continuum emission in the \textit{uv}-plane. 

The continuum-free visibilities were imaged using the \texttt{tclean} task in the ``cube'' mode using 0.1 arcsecond cells and natural weighting to create the ``dirt'' line cubes. Given that we are ultimately searching for faint emission on large scales, natural weighting is the ideal weighting scheme for us. The channel width is kept to the intrinsic value of $\sim$20\,km\,s$^{-1}$. Once we estimate the RMS of the ``dirt'' cubes, we repeat the \texttt{tclean} task, this time cleaning the emission down to $3\times$RMS. The final beam size of the natural weighted line cubes are 0.7-1.0 arcsecond. We imaged the continuum using the \texttt{tclean} task in the continuum (mfs) mode, using only the line-free channels (i.e. channels $>\pm2\times$ FWHM of the systematic redshift; which corresponds to our original assumption of line-free channels of $\pm 600$ km s$^{-1}$).

\begin{table}
\caption{Table of basic properties of our sample. (1) Object ID; (2,3) Coordinates; (4) optical redshift ; (5,6) Black hole masses and bolometric luminosities from \citealt{Hamann17}.
}
\begin{tabular}{@{}lccccc@{}} 
\hline  
\hline  
 (1) & (2) & (3) & (4) & (5) & (6) \\ 
ID & RA & Dec & z & log$_{
\rm 10}$ M$_{\rm BH}$ & log$_{
\rm 10}$ L$_{\rm bol}$\\ 
    &    &     &   &  M$_{\odot}$ & ergs s$^{-1}$\\ 
\hline  
J0006+1215&1.5445&12.2504&2.318&${10.0}$&${47.9}$ \\ 
 J0007+1222&1.9425&12.3733&2.446&${7.8}$&${47.7}$ \\ 
 J0052-0556&13.1385&-5.9482&2.363&${9.1}$&${47.4}$ \\ 
 J0826+0542&126.7226&5.7131&2.58&${8.1}$&${47.5}$ \\ 
 J0832+1615&128.0008&16.2501&2.428&${7.6}$&${47.5}$ \\ 
 J0834+0159&128.702&1.9892&2.605&${8.1}$&${47.6}$ \\ 
 J1137+1427&174.3394&14.458&2.302&${8.9}$&${48.1}$ \\ 
 J1217+0234&184.2696&2.5714&2.428&${8.5}$&${47.4}$ \\ 
 J1232+0912&188.1739&9.2026&2.405&${8.6}$&${47.8}$ \\ 
 J1316+0453&199.118&4.8878&2.16&${9.2}$&${47.9}$ \\ 
 J1342+0930&205.7269&9.5165&2.347&${8.7}$&${47.1}$ \\ 
 J1348-0250&207.0006&-2.8351&2.238&${8.5}$&${47.1}$ \\ 
 J2215-0056&333.85&-0.9455&2.508&${8.4}$&${47.1}$ \\ 
 J2223+0857&335.7797&8.9505&2.291&${8.5}$&${47.8}$ \\ 
 J2323-0100&350.8591&-1.0092&2.381&${8.4}$&${47.1}$ \\ 
 \hline  
\end{tabular} 

\label{Table:Sample}
\end{table}

\subsection{Extraction of the integrated line emission}\label{sec:spectra}

In this section, we describe the extraction and analyses of the total integrated emission lines. To get the emission-line profiles of the ERQs, we extract the spectra from the naturally weighted cubes using apertures with a radius of 0.7--0.9 arcsecond, corresponding to the beam size of these observations, centred on the peak of the emission, which we found by fitting a two-dimensional Gaussian to the emission. 

To model the extracted emission-line profiles, each emission line was fitted with one or two Gaussian components with the centroid, line width and normalisation (flux) as free parameters. We note that we do not give a physical meaning to either of these components and we use these to characterise the total emission line profile. To distinguish between individual models, several different statistics can be used. One of the most basic models is the reduced $\chi^{2}_{\rm red}$ parameter: 

\begin{equation}
\chi^{2}_{\rm red} = \frac{\chi^{2}}{N_{\rm data}- N_{\rm var}}
\end{equation}

where N$_{\rm data}$ is the number of data points and N$_{\rm var}$ is the number of variables that are fitted (e.g. three for a single gaussian profile). Alternatively, we can use the Bayesian Information Criterion (BIC; \citealt{Schwarz78}, \citealt{Liddle07}, \citealt{Concas19}), which further penalises the $\chi^{2}$ for more variables:
\begin{equation}
    \rm BIC = \chi^{2} + N_{\rm var}\log(N_{\rm data})
\end{equation}
Similarly to the $\chi^{2}_{\rm red}$, the model with lower BIC is a statistically better fit to the data. The following criteria are used: <2 - no difference; 2--6 - slight evidence; 6--10 significant evidence; >10 a better fit. However, it is worth noting that using BIC or reduced $\chi^{2}$ metrics to distinguish between models does not change the conclusions of this work. 

\begin{landscape}
\begin{figure}
    \includegraphics[width=1.1\paperwidth]{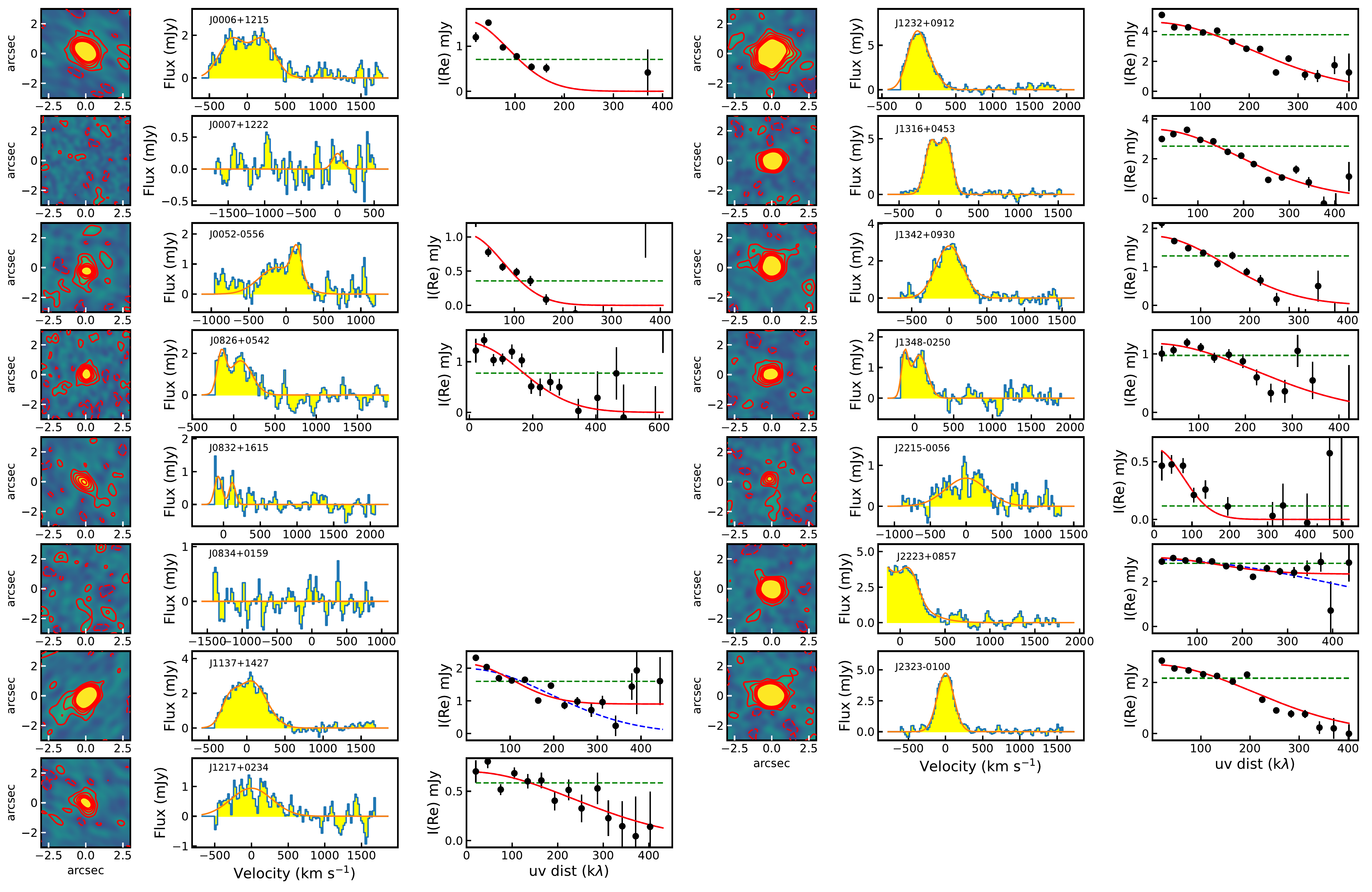}
   \caption{ Summary of the CO(7-6) data. Left column: Moment-0 map of the emission line. The red solid contours show 1, 2, 3, 4, 5 $\sigma$ and the red dashed contours -1, -2, -3, -4, -5 $\sigma$. Middle panel: Spectrum extracted from the region corresponding to the beam size. The orange line indicates the best fit to the data according to the BIC statistics. Right panel: Plot of the \textit{uv}-visibilities (flux (real component) vs the \textit{uv}-distance). The red solid line shows the best fit to the data. We have also included the single resolved source fit (blue dashed line), unresolved source model (green dashed line) and a combination of resolved and a point source (magenta dashed line). We only show the \textit{uv}-visibilities for targets detected over 10 $\sigma$.
    }
   \label{fig:CO_data}
\end{figure} 
\end{landscape}

\begin{landscape}
\begin{figure}
    \includegraphics[width=1.1\paperwidth]{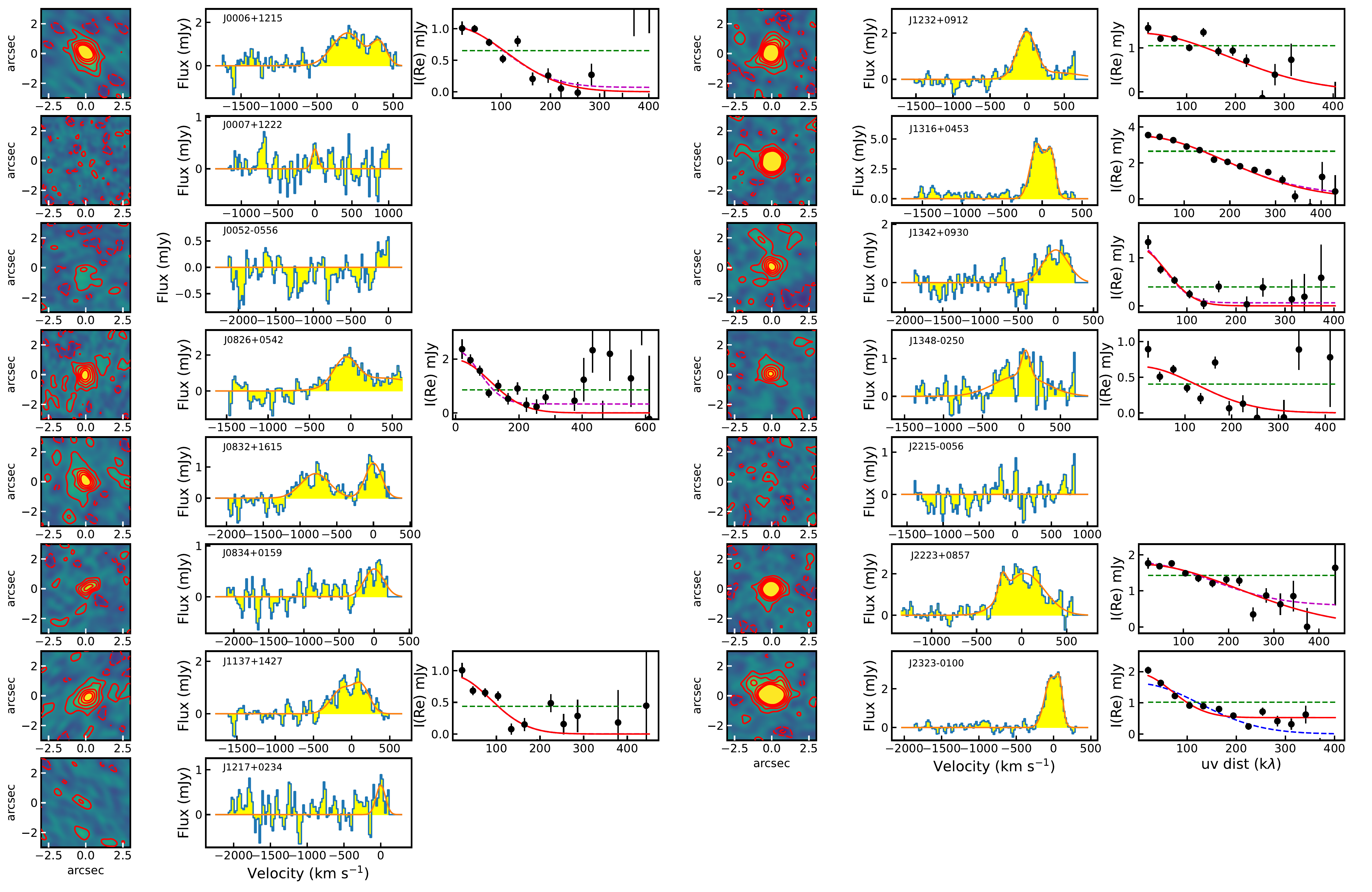}
   \caption{ Summary of the [\ion{C}{i}] data. Left column: Moment-0 map of the emission line. The red solid contours show 1, 2, 3, 4, 5 $\sigma$ and the red dashed contours -1, -2, -3, -4, -5 $\sigma$. Middle panel: Spectrum extracted from the region corresponding to the beam size. The orange line indicates the best fit to the data according to the BIC statistics. Right panel: Plot of the \textit{uv}-visibilities (flux (real component) vs the \textit{uv}-distance). The red solid line shows the best fit to the data. We have also included the single resolved source fit (blue dashed line), unresolved source model (green dashed line) and a combination of resolved and a point source (magenta dashed line). We only show the \textit{uv}-visibilities for targets detected over 10 $\sigma$.
    }
   \label{fig:CI_data}
\end{figure} 
\end{landscape}

\begin{landscape}
\begin{figure}
    \includegraphics[width=1.1\paperwidth]{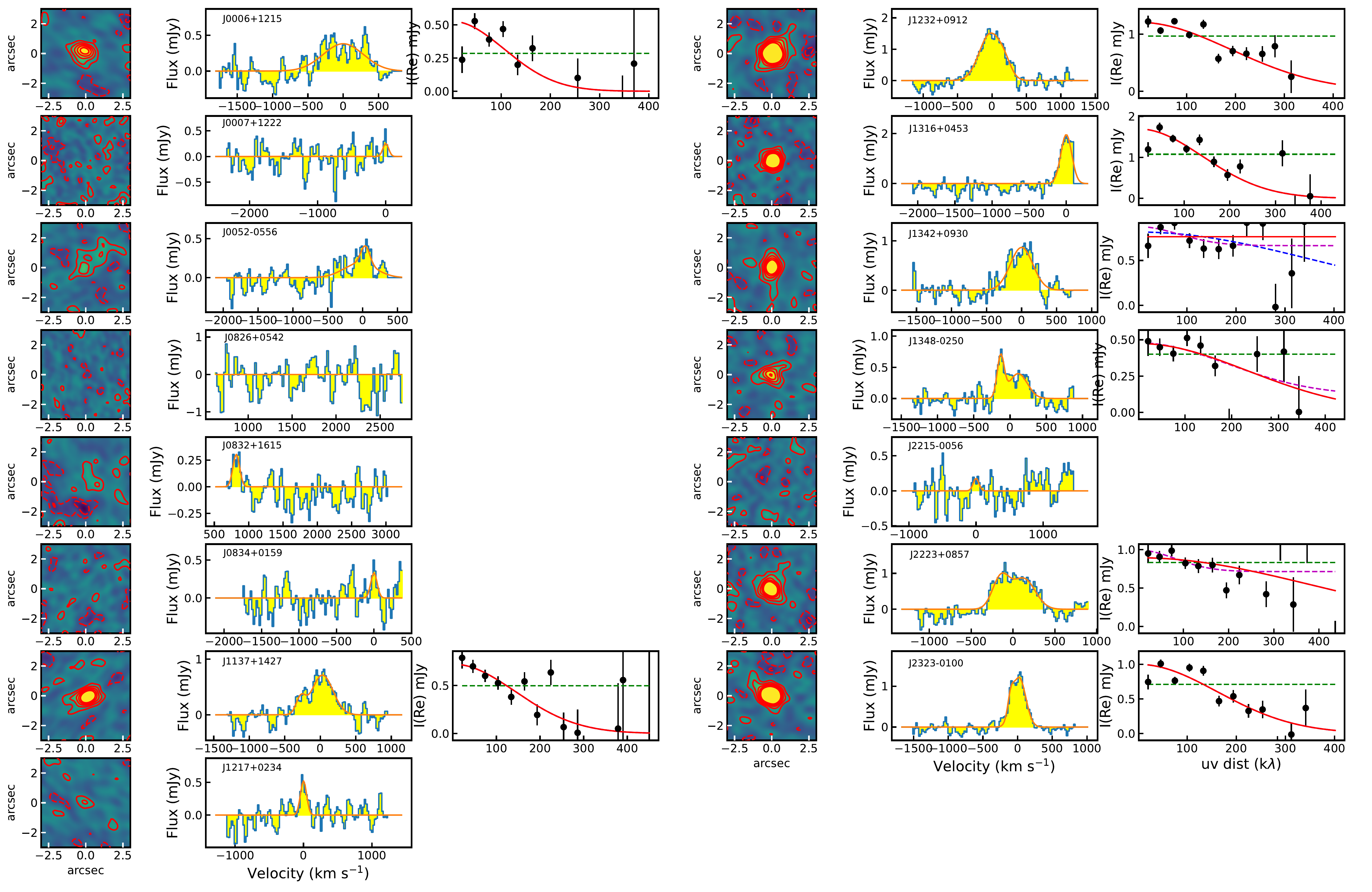}
   \caption{ Summary of the H$_2$O data.Left column: Moment-0 map of the emission line. The red solid contours show 1, 2, 3, 4, 5 $\sigma$ and the red dashed contours -1, -2, -3, -4, -5 $\sigma$. Middle panel: Spectrum extracted from the region corresponding to the beam size. The orange line indicates the best fit to the data according to the BIC statistics. Right panel: Plot of the \textit{uv}-visibilities (flux (real component) vs the \textit{uv}-distance). The red solid line shows the best fit to the data. We have also included the single resolved source fit (blue dashed line), unresolved source model (green dashed line) and a combination of resolved and a point source (magenta dashed line). We only show the \textit{uv}-visibilities for targets detected over 10 $\sigma$.
    }
   \label{fig:H2O_data}
\end{figure} 
\end{landscape}

The resulting extracted spectra with best fits are presented in Figures \ref{fig:CO_data}, \ref{fig:CI_data} \& \ref{fig:H2O_data}  and we present the results of the best-fits in Table 2. We define the FWHM of the line as the velocity width containing 68\% of the flux of the line, similar to previous works in the literature \citep[][]{Bothwell13, Wardlow18} to describe the emission linewidth in a way that is independent of the fitted model. We report SNR, integrated line flux and the velocity line width for all lines detected at SNR>5 in Table \ref{Table:Lines}.

\begin{landscape}
\begin{table}
\caption{Table of basic properties of our sample. (1) Object ID; (2) SNR of the CO(7-6) line; (3) FWHM of the CO(7-6) line; (4) redshift of the CO(7-6) line; (5) integrated flux of the CO(7-6) line; (6) SNR of the [\ion{C}{i}] line; (7) FWHM of the [\ion{C}{i}] line; (8) redshift of the [\ion{C}{i}] line; (9) integrated flux of the [\ion{C}{i}] line; (10) SNR of the H$_2$O line; (11) FWHM of the H$_2$O line; (12) redshift of the H$_2$O line; (13) integrated flux of the H$_2$O line; (14) Continuum flux at 1200 $\mu$m (observed frame).
}
\begin{tabular}{@{}lcccccccccccccc@{}} 
\hline  
\hline  
(1) & (2) & (3) & (4) & (5) & (6) & (7) & (8) & (9) & (10) & (11) &(12) & (13)  & (14)\\ 
ID & SNR$_{\rm CO(7-6)}$ & FWHM$_{\rm CO(7-6)}$ & z$_{\rm CO(7-6)}$ &  I$_{\rm CO(7-6)}$& SNR$_{\rm [CI]}$ & FWHM$_{\rm [CI]}$ & z$_{\rm [CI]}$ &  I$_{\rm [CI]}$& SNR$_{\rm H_2O}$ & FWHM$_{\rm H_2O}$ & z$_{\rm H_2O}$ & I$_{\rm H_2O}$ & S$_{1100 \mu m}$\\ 
    &    &  (km s$^{-1}$) &   & (Jy km s$^{-1}$)  & & (km s$^{-1}$) &   & (Jy km s$^{-1}$) & & (km s$^{-1}$) &   & (Jy km s$^{-1}$) & (mJy) \\ 
\hline  
J0006+1215&29.3&680$\pm 100$&2.318$\pm 0.001$&1.49$\pm 0.22$&16.6&670$\pm 100$&2.318$\pm 0.001$&0.99$\pm 0.15$&10.3&670$\pm 100$&2.318$\pm 0.001$&0.27$\pm 0.04$ & $0.87\pm0.04$ \\ 
 J0007+1222& 2.4 & - & -& - & 2.2 & - & -& - & 1.0 & - & -& -  & $<0.1$ \\ 
 J0052-0556&13.7&590$\pm 90$&2.363$\pm 0.001$&0.73$\pm 0.11$& 0.5 & - & -& - &10.2&420$\pm 60$&2.363$\pm 0.001$&0.14$\pm 0.02$ & $0.27\pm0.03$ \\ 
 J0826+0542&14.2&410$\pm 60$&2.58$\pm 0.001$&0.8$\pm 0.12$&13.0&1000$\pm 150$&2.579$\pm 0.001$&1.27$\pm 0.19$& 2 & - & -& -  & $0.87\pm0.07$ \\ 
 J0832+1615&7.7&270$\pm 40$&2.428$\pm 0.001$&0.18$\pm 0.03$&7.4&1040$\pm 160$&2.434$\pm 0.001$&0.7$\pm 0.1$& 1.2 & - & -& -  & $0.21\pm0.02$ \\ 
 J0834+0159& 0.0 & - & -& - &6.5&320$\pm 50$&2.596$\pm 0.001$&0.19$\pm 0.03$& 2.5 & - & -& -  & $0.29\pm0.05$ \\ 
 J1137+1427&59.5&560$\pm 80$&2.302$\pm 0.001$&1.62$\pm 0.24$&17.1&470$\pm 70$&2.302$\pm 0.001$&0.62$\pm 0.09$&19.4&440$\pm 70$&2.303$\pm 0.001$&0.32$\pm 0.05$ & $0.44\pm0.04$ \\ 
 J1217+0234&12.2&660$\pm 100$&2.428$\pm 0.001$&0.66$\pm 0.1$& 4.2 & - & -& - &5.5&110$\pm 20$&2.428$\pm 0.001$&0.06$\pm 0.01$ & $0.21\pm0.03$ \\ 
 J1232+0912&64.7&380$\pm 60$&2.405$\pm 0.001$&2.44$\pm 0.37$&32.3&710$\pm 110$&2.405$\pm 0.001$&0.86$\pm 0.13$&31.1&410$\pm 60$&2.405$\pm 0.001$&0.68$\pm 0.1$ & $0.69\pm0.04$ \\ 
 J1316+0453&66.7&320$\pm 50$&2.16$\pm 0.001$&1.84$\pm 0.28$&46.0&290$\pm 40$&2.16$\pm 0.001$&1.5$\pm 0.22$&20.8&170$\pm 30$&2.159$\pm 0.001$&0.35$\pm 0.05$ & $2.05\pm0.04$ \\ 
 J1342+0930&42.4&420$\pm 60$&2.347$\pm 0.001$&1.26$\pm 0.19$&11.6&390$\pm 60$&2.347$\pm 0.001$&0.47$\pm 0.07$&17.5&360$\pm 50$&2.347$\pm 0.001$&0.33$\pm 0.05$ & $0.25\pm0.05$ \\ 
 J1348-0250&18.3&320$\pm 50$&2.238$\pm 0.001$&0.5$\pm 0.08$&13.8&600$\pm 90$&2.238$\pm 0.001$&0.52$\pm 0.08$&14.7&400$\pm 60$&2.237$\pm 0.001$&0.2$\pm 0.03$ & $0.56\pm0.02$ \\ 
 J2215-0056&15.4&670$\pm 100$&2.508$\pm 0.001$&0.49$\pm 0.07$& 1.2 & - & -& - & 1.4 & - & -& -  & $0.22\pm0.05$ \\ 
 J2223+0857&42.7&1050$\pm 160$&2.291$\pm 0.001$&5.7$\pm 0.85$&31.9&560$\pm 80$&2.291$\pm 0.001$&1.22$\pm 0.18$&17.8&470$\pm 70$&2.29$\pm 0.001$&0.51$\pm 0.08$ & $1.11\pm0.03$ \\ 
 J2323-0100&64.2&260$\pm 40$&2.381$\pm 0.001$&1.25$\pm 0.19$&41.5&240$\pm 40$&2.38$\pm 0.001$&0.72$\pm 0.11$&35.0&260$\pm 40$&2.381$\pm 0.001$&0.35$\pm 0.05$ & $1.12\pm0.06$ \\ 
 \hline  
\end{tabular} 

\label{Table:Lines}
\end{table}
\end{landscape}

\section{Analysis}\label{sec:analyses}

In this section, we present the analysis used in this work to achieve our goal of tracing cold gas halos around these quasars. In \S \ref{sec:uv-analyses}, we describe our analyses of the ALMA data in the \textit{uv}-plane, and in \S \ref{sec:RP_extractio} we present the extraction and modelling of the radial brightness profiles. Finally, in \S \ref{sec:image_stack} \& \ref{sec:uv_stack} we present our method for stacking in both image and \textit{uv}-plane stacking, respectively.

\subsection{Investigating \textit{uv}-visibilities}\label{sec:uv-analyses}

In order to measure the flux and the sizes of the emission, we have extracted and collapsed the \textit{uv}-visibilities. We first split the calibrated measurement sets to include only channels with the emission line. We selected these line emission channels based on the extracted spectra in the image plane, by selecting channels with at least 10\% of the peak flux of the combined spectra. We extracted the \textit{uv}-visibilities using \texttt{uvplot} Python library \citep[][]{uvplot_tazzari} and binned them in 30\,k$\lambda$-wide bins. We show these visibilities for each emission line in Figures \ref{fig:CO_data}, \ref{fig:CI_data} \& \ref{fig:H2O_data} for sources with SNR>10. 

We fitted these amplitudes as a function of $uv$ distance with four different functions, a constant, to represent an unresolved source \citep[e.g.,][]{Rohlfs96}, a half Gaussian model centred on 0, to represent a resolved source with a Gaussian morphology, a combination of the previous two models representing both a central unresolved source and resolved source and two half Gaussians representing two resolved sources. Similarly to the spectral line fitting, we used BIC to distinguish between the models. The final uncertainties on the sizes comes as 68\% confidence interval uncertainties from the MCMC fitting. We summarise the estimated sizes in Table \ref{Table:Sizes}.

\subsection{Extraction and modelling of radial profiles}\label{sec:RP_extractio}

One of the goals of this work is to search for extended molecular gas emission (in CO(7-6), [\ion{C}{i}], H$_2$O and dust continuum). To achieve this, we construct the radial brightness profiles of the moment-0 map, i.e. the flux map. Compared to curve-of-growth, radial brightness profiles are more sensitive to faint emission at larger radii - i.e. extended halo emission. Although our sources were the primary targets of our ALMA observations, the emission line regions are not in the exact center of the field (1-2 pixels offset). Hence, we fitted the objects with a 2D Gaussian profile to find the centre of the emission. We then calculated the median flux in annuli with a width of 0.2 arcsecond (2 pixels), with the annuli centred on the coordinates determined from the 2D Gaussian fitting. The uncertainty on the flux is calculated as the RMS of the moment-0 map divided by the square root of the number of beams covered by the annuli (the number of pixels in the annuli divided by the number of pixels in the beam). We estimated the RMS of the noise of the moment-0 map using the \texttt{astropy}'s sigma-clipping function (SNR=3). We have repeated this procedure on an image of the clean elliptical beam to compare the radial profiles of the object to the radial profiles of the beam. 

We investigated the effect of the width of the annuli versus the number of the annuli on the radial profile modelling. As we increase the width of the annuli the uncertainty on the median flux decreases. However, the increased size of annuli results in fewer annuli, to avoid overlap of the bins. Comparing the $\Delta$BIC between the models (see below) showed that the decrease of the flux uncertainties with increased annulus width does not out-weight the downside of more coarse sampling of the radial surface brightness profile. That is, in this case, we found that having more data points with large uncertainties is better than having a smaller number of data points with smaller uncertainties. 

\subsubsection{Modelling of the radial brightness profiles}\label{sec:RP_analyses}

In order to investigate whether the data contains a single galaxy component (i.e. the gas in the galaxy) or whether it also contains a diffuse outer larger-scale component (i.e. a halo), we modelled radial surface density profiles of a single symmetrical 2D Gaussian or two symmetrical 2D Gaussian components. 
%Throughout this work, we will define any additional emission component, significantly larger than the galaxy component as a halo. 

We start by creating a 2D circular Gaussian model with arbitrary amplitude (=1) and set the intrinsic size of the object (FWHM). We convolved the model with the beam and extracted the radial brightness profile of this mock convolved image as described in \S~\ref{sec:RP_extractio}. Finally, we compared the modelled and observed radial brightness profiles and used \texttt{Python}'s \texttt{emcee} library to find the best value of FWHM using the MCMC ensemble sampler algorithm. 
We set a flat prior on the FWHM to be between 0 and 3 arcseconds. For the double 2D Gaussian model, we fit three free parameters: FWHM$_{\rm galaxy}$, FWHM$_{\rm halo}$ and the log$_{10}$ ratio between the peak flux of the galaxy and outer components log$_{10}\frac{\rm peak_{\rm halo}}{\rm peak_{\rm galaxy}}$. For these three parameters we set the priors to [0, 1 arcsec], [1, 4 arcsec] and [-1, -5] for FWHM$_{\rm galaxy}$, FWHM$_{\rm halo}$ and log$_{10}\frac{\rm peak_{\rm halo}}{\rm peak_{\rm galaxy}}$, respectively. We adopt the likelihood function of \citet{Nikolic09} as:
\begin{equation}
  \log L(\sigma) = \sum_{i} \left[ \left[ \frac{D_i- M_i}{\delta D_i}\right]^2 +\log(2\pi\delta D_i^2)\right]
\end{equation}
where the D$_{i}$ and $\delta$D$_{i}$ are the data and uncertainties on the data, respectively, M$_{i}$ is the model. The final results quoted in this work are the 50th, 16th and 84th percentiles of the posterior distribution. 

\subsection{Stacking in the image plane}\label{sec:image_stack}

Although we detect a majority of the sources in each observed emission line, we also stack the emission line cubes to search for the faint emission on large scales. We stacked the individual cubes rather than moment-0 maps of the emission lines. There is no reason to assume that the line widths of the extended and galaxy component are the same, which would result in removing some extended component signal. Stacking the cube rather than moment-0 maps allows us to search for the extended halo emission across multiple velocity ranges later on. Furthermore, we only stacked cubes which contain a detected emission line. As we described above, there can be a significant offset between the optical and submm lines and which would potentially result in including noise only. 

To stack the emission cubes, we used the same code from \citet{Jones22}, a similar technique to \citet{Delhaize13,Bischetti19, Jolly20}. Here we briefly outline the method. To stack the emission line cubes, we first start with an empty cube with a spatial dimension of 128$\times$128 pixels (corresponding to 12.8$\times$ 12.8 arcseconds) with a spectral axis of -2000--2000 km s$^{-1}$ in 200 channels, resulting in channels width of 20 km s$^{-1}$. This stacked cube setup was chosen to match the individual cubes from the imaging pipeline. For each of the emission line cubes, we created a cutout corresponding to the size of the new empty cube. However, the spectral scale is different for each cube as it is tuned to the different central frequencies and hence velocity width. As a result, we distribute the flux to individual velocity bins of the stacked cube as described by equation 1 in \citet{Jones22}.

We stacked the cubes based on four different weighting schemes: 
\begin{enumerate}
    \item \textit{Uniform:} The cubes are stacked without any weighting schemes.
    
    \item \textit{Normalisation (Normed):} We estimated the maximum value of each cube and calculated the weighting as 1/(maximum value). This weighting scheme effectively results in normalising the data cubes by their maximum values.
    
    \item \textit{Inverse Variance (InvV):} Normalising by the noise levels in the cube. We first estimated the RMS of each cube using sigma clipping of SNR=3. We then calculated the weights as 1/(RMS$^{2}$), which penalises data cubes with a larger noise. However, given the uniformity of the depth of our observations, the weights are very similar resulting to similar results to the uniform weighting.
    
    \item \textit{Inverse Variance \& Normalization (InNo):} Combination of the previous two weighting schemes, with the final weights being 1/(RMS$^{2}\times$ max value).
\end{enumerate}

To accurately describe the radial emission line profiles, we need to also stack the beams of the individual cubes to find the final beam of the stacked cube. As each spectral window has a narrow frequency range, the beam size is changing only by maximum of $\sim 0.5$\%. As this difference is negligible, we use only a single common beam for the cube to simplify the procedure and the analysis of the stacked cubes. Using a single common beam per data cube simplifies the beam stacking and data analysis from three dimensions to two dimensions. 
We extracted the clean beam information from the header of the individual cube and created an image of the beam for each cube. We stacked these beam images on the same grid as the stacked cube (128$\times$128 pixels corresponding to 12.8$\times$ 12.8 arcsec), giving the beam the same weight as to the individual cube in the cube stacking. We then fit this stacked beam image with a 2D Gaussian, similarly to the method used by the \texttt{tclean} to find the size and shape of the clean beam. This then became the new clean beam for the stacked data. 

Using the method described above, we stacked the data cubes of CO(7-6), [\ion{C}{i}], H$_2$O and continuum emission. We constructed moment-0 maps of stacked cubes for each of the weightings in the range of $\pm$100 km s$^{-1}$, $\pm$200 km s$^{-1}$, $\pm$300 km s$^{-1}$, $\pm$400 km s$^{-1}$ and $\pm$500 km s$^{-1}$. For CO(7-6), the emission line is at the edge of the band in seven objects, with five objects with CO(7-6) less than 500 km s$^{-1}$ from the edge of the band. However, as we are not measuring the emission line profiles of the stacked data this does not influence our results; although it will decrease the SNR of our stacked data. We showed these moment-0 maps and their extracted brightness profiles in Figures A\ref{fig:CO_data}, A\ref{fig:CI_data}, A\ref{fig:H2O_data}. 

\subsection{Stacking in \textit{uv}-plane}\label{sec:uv_stack}

Since our data was taken with an interferometer, it is important to verify any morphology results by investigating the data in the \textit{uv}-plane. The image analysis from interferometric observations can be sensitive to the exact cleaning procedure, which can accidentally introduce faint artefacts into the data. Furthermore, any extended emission detected in the image stacks can be caused by stacking the residuals of the dirty beam. Fortunately, working in the \textit{uv}-plane circumvents all of these downsides of image-plane stacking. 

For each object, we first split the visibility data into separate measurement sets, containing: CO(7-6), [\ion{C}{i}], H$_2$O and continuum emission only. For the emission line visibilities, we extracted channels in the velocity range as the stacked cube with strongest halo emission from the image-based stacking: $\pm$300, $\pm$200 and $\pm$200 kms$^{-1}$ for CO(7-6), [\ion{C}{i}] and H$_2$O, respectively (see \S \ref{sec:image_stack_res}). We centred velocity ranges on the frequency peak (the 3D position in our ALMA cubes. For the continuum, we selected line-free channels (defined as 2$\times$ FWHM from the spectral fitting). During the \texttt{split} task, we also binned the data in the time domain with 30s bins, to make the data sizes more manageable. We verified that the time-binning does not affect our final results, by stacking a subset of the [\ion{C}{i}] data set.

In order to stack the visibilities, we used the \texttt{STACKER} \citep[][]{Lindroos2015}. We first concatenate (using the CASA \texttt{concat} task) the individual emission line and continuum measurement sets to create a single measurement set per emission tracer. We shifted the coordinate of the visibility data sets by rewriting the source centre determined from the moment-0 maps as “00:00:00.00, 00:00:00.0” using the \texttt{uv.stack} task in \texttt{STACKER}. Finally, we recalculated the data weights for the combined visibility data sets with the \texttt{statwt} task, based on the scatter of visibilities, which includes the effects of integration time, channel width, and system temperature. This is comparable to using the inverse rms weighting for the image-based stacking.

We have adapted our \textit{uv}-stacking procedure for the continuum stacking. Both \texttt{stacker} and \texttt{uvplot} require for all the spectral windows to have the same number of channels. This is not a problem for the emission line stacking, as we always stack the same number of channels (i.e. same line width). However, for the continuum, each spw has a different number of line-free channels. The above-outlined process would result in a significant number of continuum channels being excluded as we would have to settle for the lowest common number of channels across thirteen objects with four spw each. Instead, for the continuum emission, we shifted the coordinate of each MS containing the continuum emission only using the \texttt{uv.stacker}, extracted the visibilities using \texttt{uvplot} and then concatenated the text files containing each of the objects \textit{uv}-visibilities. This is the equivalent of stacking the objects in a uniform scheme, as we cannot use the \texttt{statwt} on the already extracted $uv$-visibilities. 

We analyse the stacked visibility datasets in the same way as visibilities of the individual objects. %As the stacked measurement sets have a spectral setup from each individual object, we cannot image the stacked data as a cube, as it would create a cube with the individual emission-line observations. Instead, we image the data in the "mfs" spectral mode (continuum mode) to create a collapsed line emission image (essentially a moment-0 map). W
We extracted the visibilities and collapsed the visibilities as described in \S \ref{sec:uv-analyses}, with bin sizes of 20 k$\lambda$. We present the stacked visibilities in Figure \ref{fig:uv_stacking}.

\section{Results \& Discussion}\label{sec:results}

In this section, we present the results of our data analyses described above. We show the overview of the emission lines in \S \ref{sec:em-lines}, sizes of the different cold gas tracers in \S~\ref{sec:galaxy-sizes} and investigate the stacked data for presence of cold gas halos in \S~\ref{sec:image_stack_res} \& \ref{sec:uv_stack_res}. In \S~\ref{sec:mass_estimates} we derived the physical properties of these halos and in \S~\ref{sec:origins} we discuss their origins.

\subsection{Individual quasar host galaxies}\label{sec:em-lines} 

First, we give a brief overview of the detected emission line properties of our sample. We detect twelve objects in CO(7-6), eleven in  [\ion{C}{i}] emission, ten objects in H$_2$O emission and fourteen objects in continuum, across a range of SNR (3--67). The measured FWHM of the emission lines ranged 220--1060 km s$^{-1}$. Although some objects require two Gaussian components to accurately describe the emission line profiles, we do not detect any evidence for broad wings indicating a large scale outflow. For objects which were detected in multiple emission lines, the line widths of different emission lines agree within the 1 sigma errors. We summarise the redshifts, SNRs, FWHM and integrated flux in Table \ref{Table:Lines}.

\subsubsection{Galaxy component sizes}\label{sec:galaxy-sizes}

We used the collapsed \textit{uv}-visibilities to measure the sizes of the CO(7-6), [\ion{C}{i}], H$_2$O and continuum in \S \ref{sec:uv-analyses} and we present the collapsed visibilities in the right panels of Figures \ref{fig:CO_data}, \ref{fig:CI_data}, \ref{fig:H2O_data}. We investigate the \textit{uv}-visibilities of all emissions above SNR=10, as this is considered a minimal SNR limit to reliably measure sizes for interferometric data \citep[see][for details]{Simpson15,Harrison16Alm, Scholtz20}. We resolve eleven out of twelve sources in CO(7-6), nine out of ten in [\ion{C}{i}] and seven out of eight sources in H$_2$O emission. For objects J1348-0250 and J2323-0100, we are able to decompose the emission into a point source and a resolved component in CO(7-6) and [\ion{C}{i}] emission respectively. In these cases, we report the values of the resolved components and mark each source with an asterisk in Table~\ref{Table:Sizes}.

\begin{figure}
    \includegraphics[width=0.9\columnwidth]{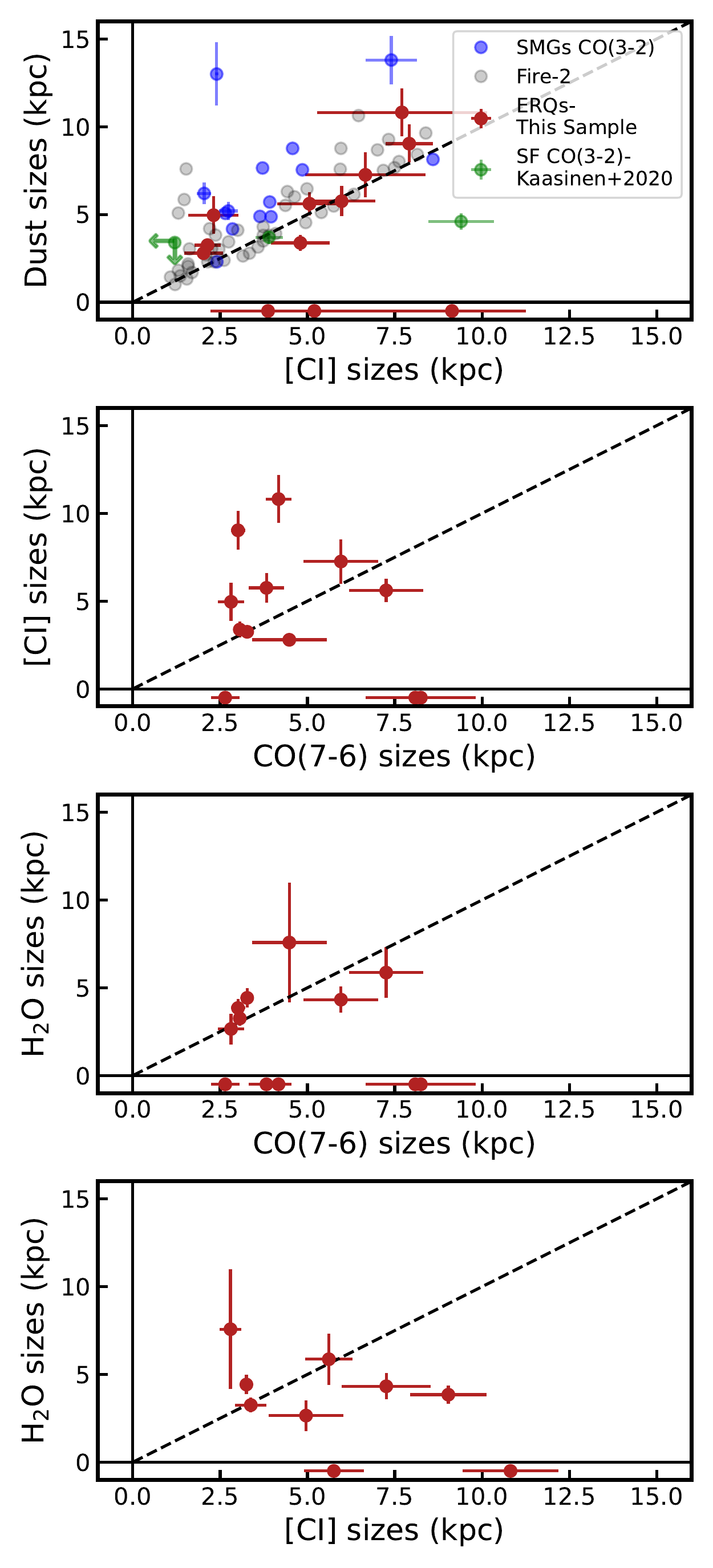}
   \caption{ Comparison of sizes from different cold gas tracers (CO(7-6), [\ion{C}{i}](2-1), H$_2$O) and dust continuum. In each panel, the dashed black line shows the 1:1 ratio between the sizes. If a source is not detected in one tracer, we only show the size for the other tracer. Points with a size of -0.5 kpc indicate an upper limit on the size in that tracer. For comparison of dust continuum vs [\ion{C}{i}](2-1), we compared our sample to other studies which measured CO(3-2) or lower, as they trace molecular gas of same temperature and density.
   }
   \label{fig:Size_comp}
\end{figure} 

We present the comparison of the CO(7-6),  [\ion{C}{i}], H$_2$O and dust continuum sizes in Figure \ref{fig:Size_comp}. We measured the sizes of the four emission tracers to be in the range of 3.1--8.6 kpc (median of $3.8\pm 2.0$ kpc) for CO(7-6), 2.9--11.2 kpc (median of $5.3\pm2.5$ kpc) for [\ion{C}{i}], 2.7--6.0 kpc (median of $4.3\pm 1.5$ kpc) for H$_2$O and 2.1-9.5 kpc (median of $5.1\pm2.6$) for FIR emission. The error on the median was estimated as a standard deviation. We summarise the galaxy sizes in each tracer in Table \ref{Table:Sizes}. There is no evidence for the evolution of either dust continuum or [\ion{C}{i}] sizes as a function of QSO bolometric luminosity, however, this can be due to covering a very narrow range of bolometric luminosities in this sample. 

In the top panel of Figure \ref{fig:Size_comp}, we compare the sizes of the dust emission with [\ion{C}{i}] for our objects (red points), SMGs \citep[blue points;][]{Chen17, Tadaki17}, star-forming galaxies \citep[green points;][]{Kaasinen20} and results from the FIRE-2 simulations \citep[grey points;][]{Cochrane19}. For the SMGs and star-forming galaxies, we use the observations of CO(3-2) emission, as [\ion{C}{i}](2-1) is rarely detected let alone resolved in high redshift galaxies. The cold molecular gas and dust sizes lie very close to the dashed 1:1 line, indicating very similar sizes.  

The CO(7-6), [\ion{C}{i}] and H$_2$O sizes are consistent with CO sizes measured by \citet{Chen17, Callistro-Rivera18} and [\ion{C}{ii}] sizes measured by ALPINE survey \citep[][]{Fujimoto20} and Hot Dust Obscured Galaxies (Hot DOGs; Scholtz et al. in prep). The median FIR sizes of our sample agree with those found in AGN host galaxies \citep[][]{Harrison16Alm, Scholtz20, Lamperti21, Scholtz21}, Hot DOGs (Scholtz et al., in prep) and sub--mm and star-forming galaxies \citep[e.g.][]{Ikarashi15, Simpson15,Hodge16, Spilker16, Tadaki17,Fujimoto18, Lang19, Chen20}, however, the range of the values is a factor of $\sim$1.5 higher than those found in submm and AGN host galaxies. This can support the hypothesis that these objects are in a blowout phase of galaxy evolution.

\begin{table}
\caption{Sizes of the different emission region from the \textit{uv}-visibilities (see \S \ref{sec:uv-analyses}). (1) Object ID; (2) Size of the CO(7-6) emission; (3) Size of the [\ion{C}{i}] emission; (4) Size of the H$_2$O emission; (5) Size of the dust continuum.
}
\begin{tabular}{@{}lcccc@{}} 
\hline  
\hline  
(1) & (2) & (3) & (4) & (5)\\ 
ID & FWHM$_{\rm CO(7-6)}$ & FWHM$_{\rm [CI]}$ & FWHM$_{\rm H_2 O}$ & FWHM$_{\rm cont}$\\ 
    &  (kpc) & (kpc) & (kpc) & (kpc) \\ 
\hline  
J0006+1215& $7.5\pm1.1$ & $5.8\pm0.7$ & $6.1\pm1.5$ & $5.2\pm1.0$  \\ 
 J0007+1222& -& -& -& - \\ 
 J0052-0556& $8.4\pm1.1$ & -& -& $4.0\pm1.7$  \\ 
 J0826+0542& $4.0\pm0.5$ & $6.0\pm0.9$ & -& $6.2\pm1.0$  \\ 
 J0832+1615& -& -& -& $5.4\pm1.9$  \\ 
 J0834+0159& -& -& -& $9.5\pm2.2$  \\ 
 J1137+1427& $6.2\pm1.1$ $^*$ & $7.5\pm1.3$ & $4.5\pm0.8$ & $6.9\pm1.8$  \\ 
 J1217+0234& $2.7\pm0.4$ & -& -& $0.0\pm0.0$  \\ 
 J1232+0912& $3.2\pm0.2$ & $3.5\pm0.5$ & $3.4\pm0.4$ & $5.0\pm0.9$  \\ 
 J1316+0453& $3.4\pm0.2$ & $3.4\pm0.1$ & $4.6\pm0.6$ & $2.2\pm0.4$  \\ 
 J1342+0930& $4.3\pm0.4$ & $11.2\pm1.4$ & -& $8.0\pm2.5$  \\ 
 J1348-0250& $2.9\pm0.4$ & $5.1\pm1.1$ & $2.8\pm0.9$ & $2.4\pm0.7$  \\ 
 J2215-0056& $8.6\pm1.6$ $^*$ & -& -& $0.0\pm0.0$  \\ 
 J2223+0857& $4.6\pm1.1$ & $2.9\pm0.3$ & $7.9\pm3.5$ & $2.1\pm0.6$  \\ 
 J2323-0100& $3.1\pm0.2$ & $9.4\pm1.1$ $^*$ & $4.0\pm0.5$ & $8.2\pm0.7$  \\ 
 \hline  
\end{tabular} 

\par $^*$ Object has two components: a point source and a resolved component. 
\label{Table:Sizes}
\end{table}

\subsubsection{Detection of cold gas halos in individual sources}\label{sec:halos_ind}

\begin{figure}
    \includegraphics[width=0.8\columnwidth]{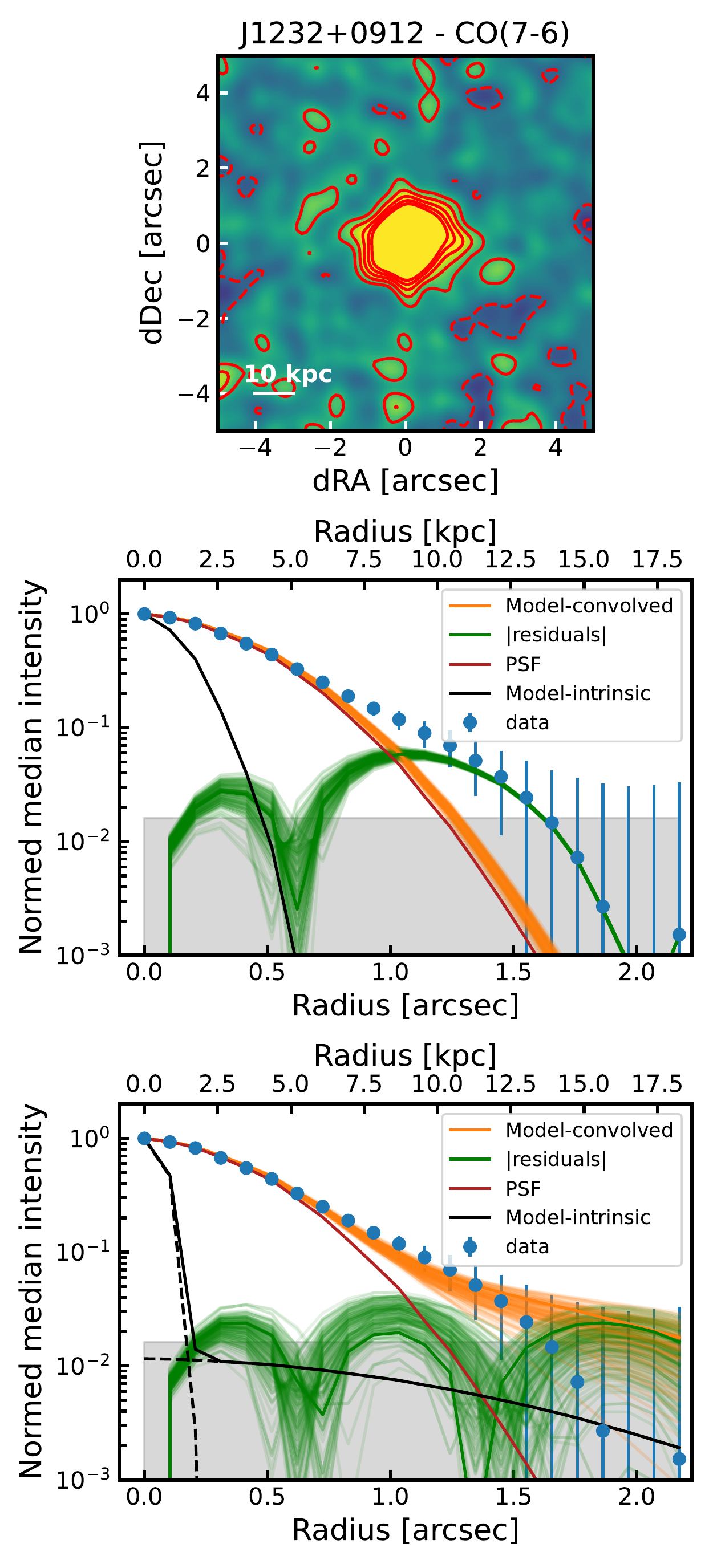}
   \caption{ Modelling of the aperture profiles of the CO(7-6) emission from J1232+0912. Top panel: The red solid contours show 1, 2, 3, 4, 5 $\sigma$ and the red dashed contours -1, -2, -3, -4, -5 $\sigma$. Middle and bottom panels: Modelling of the radial brightness profiles using a single resolved source model (middle panel) and two resolved sources model. The blue points show the extracted radial brightness profile and their uncertainties. The orange and green lines show 100 randomly drawn solutions from the MCMC chain for the fit and residuals, respectively. The black line shows the intrinsic model before the convolution. The shaded region shows the 0.5$\times$ RMS of the moment-0 maps.
    }
   \label{fig:123241_CO_Aperture_modelling}
\end{figure}

\begin{figure}
    \includegraphics[width=0.8\columnwidth]{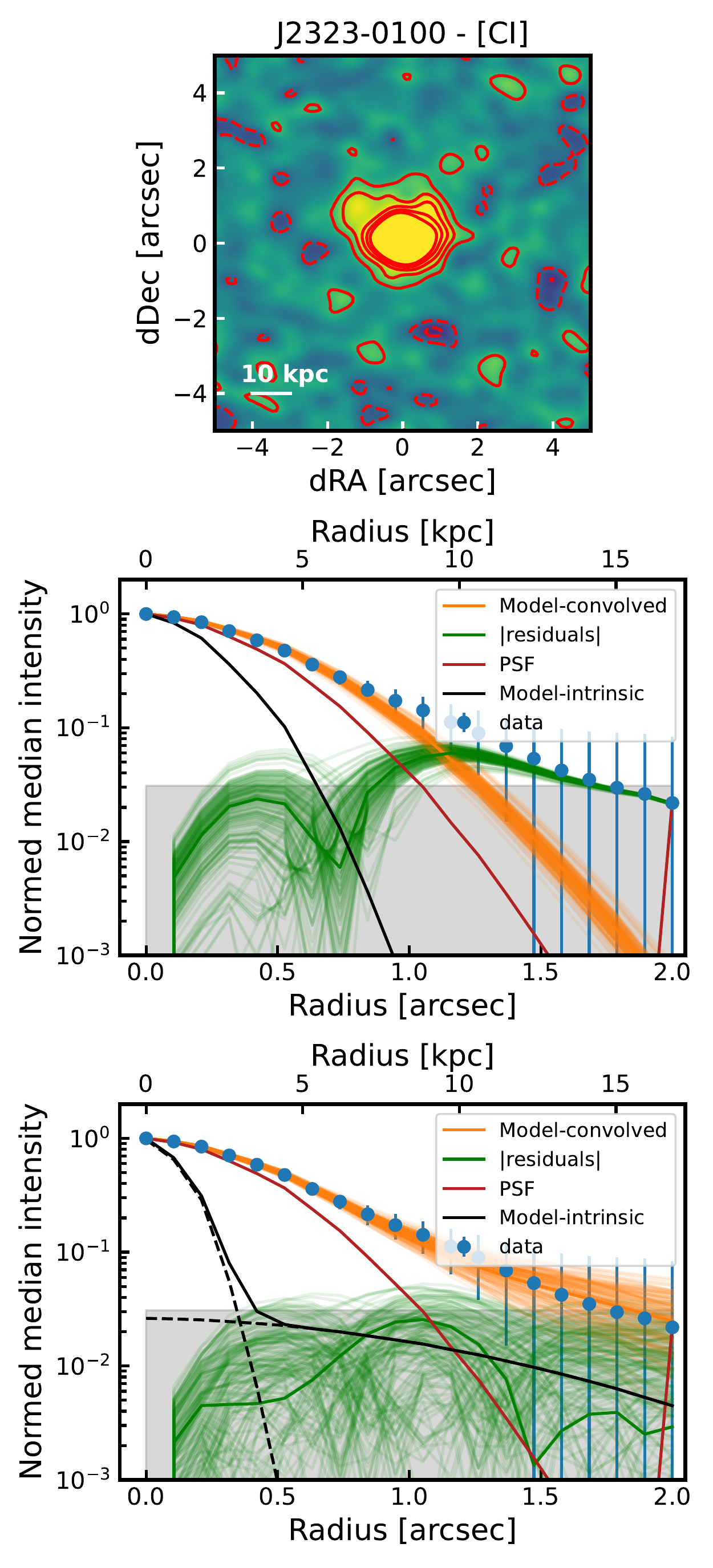}
   \caption{ Modelling of the aperture profiles of the [\ion{C}{i}] emission from J2323-0100. Top panel: The red solid contours show 1, 2, 3, 4, 5 $\sigma$ and the red dashed contours -1, -2, -3, -4, -5 $\sigma$. Middle and bottom panels: Modelling of the radial brightness profiles using a single resolved source model (middle panel) and two resolved sources model. The blue points show the extracted radial brightness profile and their uncertainties. The orange and green lines show 100 randomly drawn solutions from the MCMC chain for the fit and residuals, respectively. The black line shows the intrinsic model before the convolution. The shaded region shows the 0.5$\times$ RMS of the moment-0 maps.
    }
   \label{fig:232326_CI_Aperture_modelling}
\end{figure}

We extracted the radial brightness profiles for our sources in CO(7-6), [\ion{C}{i}] and H$_2$O emission in \S~\ref{sec:RP_extractio} and we modelled these radial brightness profiles as described in \S~\ref{sec:RP_analyses}. Overall we detect an additional large-scale extended (halo) emission in two sources: J1232+0912 (CO 7-6) and J2323-0100 ([\ion{C}{i}]) and we show these in Figures \ref{fig:123241_CO_Aperture_modelling} \& \ref{fig:232326_CI_Aperture_modelling}. We show the results of the fitting of a single resolved component in the middle panels of Figures \ref{fig:123241_CO_Aperture_modelling} \& \ref{fig:232326_CI_Aperture_modelling}, the residuals (green lines) showing a clear need for an additional large scale halo component. We present the results of fitting two resolved components - galaxy component and a large scale halo component - in the bottom panels of Figures \ref{fig:123241_CO_Aperture_modelling} \& \ref{fig:232326_CI_Aperture_modelling}. Based on the BIC and reduced $\chi^2$, the radial brightness profiles require both components to be fitted. The FWHM size of the large-scale extended halo emission extended emission is $ 22.38 ^{+ 5.26 }_{- 4.09 }$ and $ 20.5 ^{+ 7.53 }_{- 5.32 }$\,kpc for J1232+0912 (CO 7-6) and J2323-0100 ([\ion{C}{i}]), respectively. The $\Delta$BIC for J1232+0912 (CO 7-6) and J2323-0100 are -24.0 and -4.0 in favour of a double component fit, respectively. We present the sizes of the individual components, BIC values and ratio between the peaks of the components in Table: \ref{Table:halo_solo}. This is a first detection of large-scale gas reservoirs around individual quasar host galaxies using the [CI] \& CO(7-6) emission line at high redshift. 

\begin{table*}
\caption{Radial surface brightness profile fitting results for objects with detected extended emission: J1232+0912 and J2323-0100. (1) Object ID; (2) Tracer in which we detect the extended emission; (3) Model fitted to the data (best fit model is in bold); (4) Size of the galaxy component; (5) Size of the outer component; (6) log$_{10}$ ratio between the peaks of the two-component; (7) BIC of the fit; (8) $\chi^2$ of the fit. }
\begin{tabular}{@{}lccccccc@{}} 
\hline  
\hline  
(1) & (2) & (3) & (4) & (5) & (6) & (7)& (8)\\ 
ID &Tracer & Model & FWHM(galaxy) & FWHM(halo) & log$_{10}$($\frac{\rm Peak_{\rm outer}}{\rm Peak_{\rm galaxy}}$) & BIC & $\chi^2$\\ 
    &     &  & [arcsec]& [arcsec]& & \\
\hline  

\multirow{2}{*}{J1223+0912}  & \multirow{2}{*}{CO(7-6)} & Single & $ 0.35 ^{+ 0.03 }_{- 0.04 }$ & - & - &47.8& 44\\
                  &       & \textbf{Double} & $ 0.07 ^{+ 0.06 }_{- 0.05 }$ & $ 2.7 ^{+ 0.63 }_{- 0.49 }$  &  $ -3.04 ^{+ 0.22 }_{- 0.13 }$ & 24.3 & 15\\ [2pt]
\hline                          
\multirow{2}{*}{J2323-0100} &\multirow{2}{*}{[CI]}    & Single & $ 0.56 ^{+ 0.05 }_{- 0.05 }$ & - & - & 13.2& 10 \\
                 &        & \textbf{Double} &  $ 0.25 ^{+ 0.15 }_{- 0.17 }$ & $ 2.47 ^{+ 0.91 }_{- 0.64 }$  & $ -2.06 ^{+ 0.26 }_{- 0.47 }$ & 9.2&1.4\\ [2pt]
\hline
\end{tabular} 
\label{Table:halo_solo}
\end{table*}

\subsection{Stacking results}

In this section we describe the results obtained from stacking the data in both image-plane and the \textit{uv}-plane. We described the stacking methods in \S \ref{sec:image_stack} \& \ref{sec:uv_stack}.

\subsubsection{Image-plane stacking results}\label{sec:image_stack_res}

We show the full results of the image-stacked data for CO(7-6), [\ion{C}{i}], H$_2$O and continuum in Figures \ref{fig:Stacking_CO}, \ref{fig:Stacking_[CI]}, \ref{fig:Stacking_H2O]}, \ref{fig:Stacking_cont]}, showing both the moment-0 map and extracted radial surface brightness profiles for each of the velocity ranges ($\pm100$, $\pm200$, $\pm 300$,$\pm 400$ and $\pm500$ km s$^{-1}$) and stacking weighting schemes. For further analysis, we use the inverse RMS weighting scheme, as it allows more direct comparison with \textit{uv}-plane stacking. Since we do not expect any \textit{a priori} correlation between the galaxy and halo emission, we do not bias the stacks based on the galaxy gas brightness such as in the Normed and InNo weighting schemes. 

\begin{figure}
    \includegraphics[width=0.8\columnwidth]{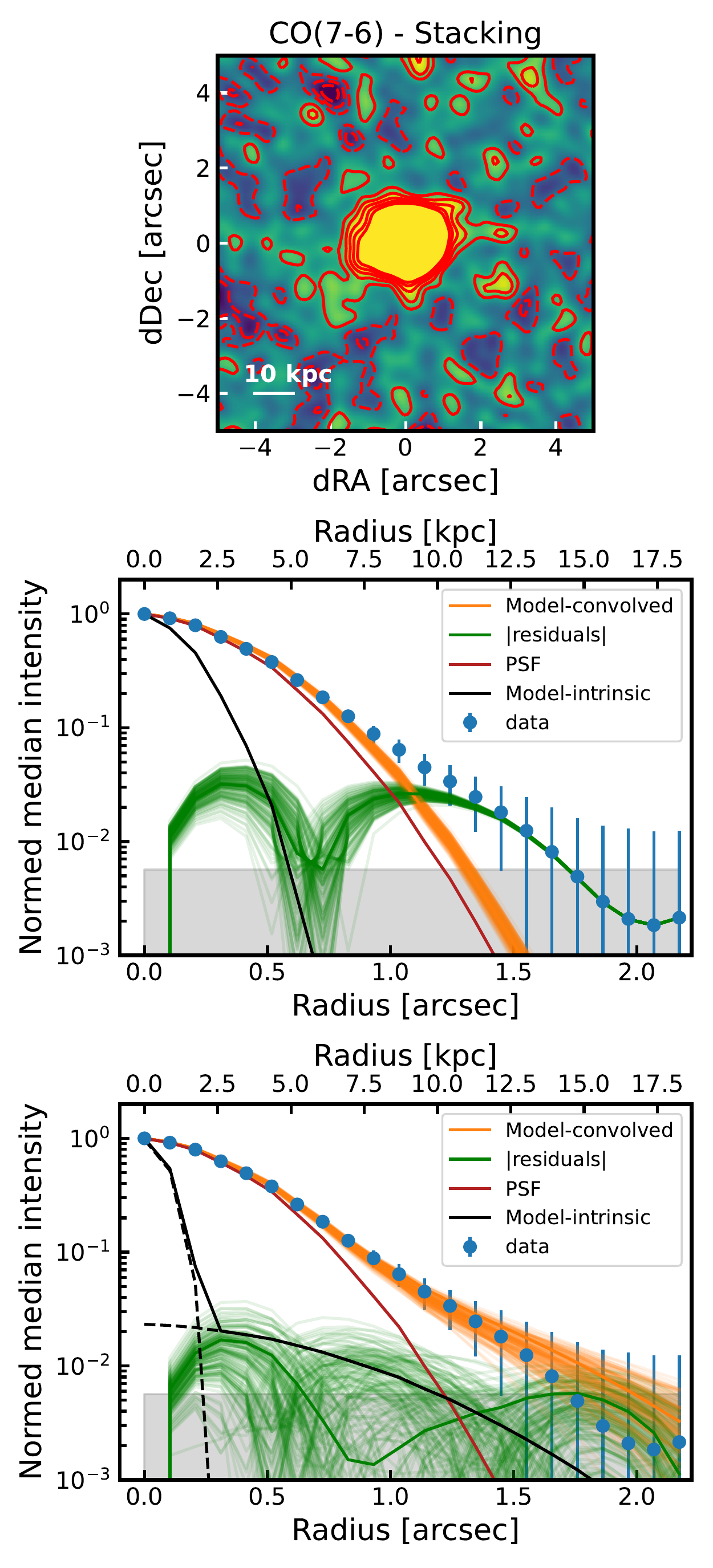}
   \caption{ Modelling of the aperture profiles of the CO(7-6) stacked emission. Top panel: The red solid contours show 1, 2, 3, 4, 5 $\sigma$ and the red dashed contours -1, -2, -3, -4, -5 $\sigma$. Middle and bottom panels: Modelling of the radial brightness profiles using a single resolved source model (middle panel) and two resolved sources model. The blue points show the extracted radial brightness profile and their uncertainties. The orange and green lines show 100 randomly drawn solutions from the MCMC chain for the fit and residuals, respectively. The black line shows the intrinsic model before the convolution. 
    }
   \label{fig:CO_Aperture_modelling}
\end{figure}

\begin{figure}
    \includegraphics[width=0.8\columnwidth]{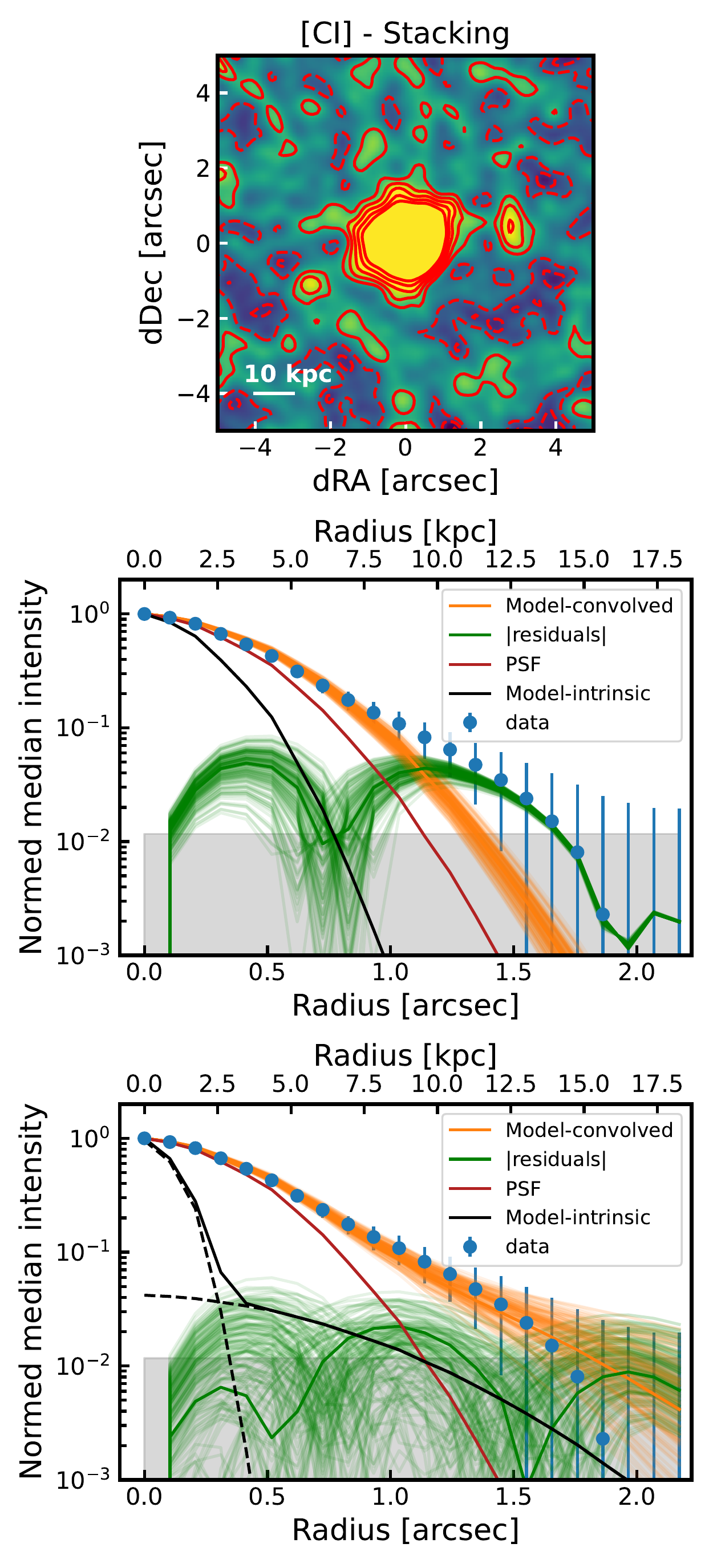}
   \caption{ Modelling of the aperture profiles of the [\ion{C}{i}](2-1) stacked emission. Top panel: The red solid contours show 1, 2, 3, 4, 5 $\sigma$ and the red dashed contours -1, -2, -3, -4, -5 $\sigma$. Middle and bottom panels: Modelling of the radial brightness profiles using a single resolved source model (middle panel) and two resolved sources model. The blue points show the extracted radial brightness profile and their uncertainties. The orange and green lines show 100 randomly drawn solutions from the MCMC chain for the fit and residuals, respectively. The black line shows the intrinsic model before the convolution.  The shaded region shows the 0.5$\times$ RMS of the moment-0 maps.
    }
   \label{fig:CI_Aperture_modelling}
\end{figure}

\begin{figure}
    \includegraphics[width=0.8\columnwidth]{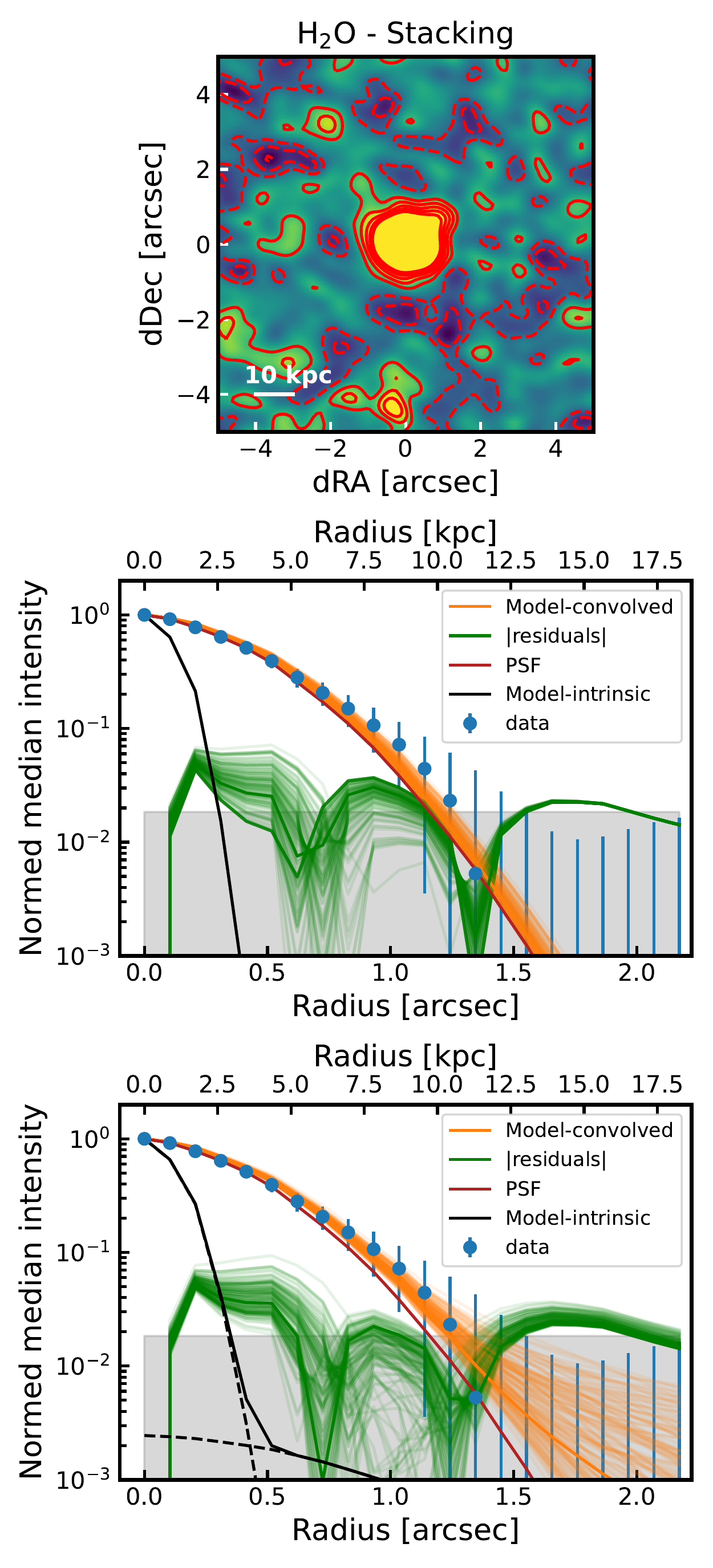}
   \caption{ Modelling of the aperture profiles of the H$_{2}$O stacked emission. Top panel: The red solid contours show 1, 2, 3, 4, 5 $\sigma$ and the red dashed contours -1, -2, -3, -4, -5 $\sigma$. Middle and bottom panels: Modelling of the radial brightness profiles using a single resolved source model (middle panel) and two resolved sources model. The blue points show the extracted radial brightness profile and their uncertainties. The orange and green lines show 100 randomly drawn solutions from the MCMC chain for the fit and residuals, respectively. The black line shows the intrinsic model before the convolution.  The shaded region shows the 0.5$\times$ RMS of the moment-0 maps.
    }
   \label{fig:H2O_Aperture_modelling}
\end{figure}

\begin{figure}
    \includegraphics[width=0.8\columnwidth]{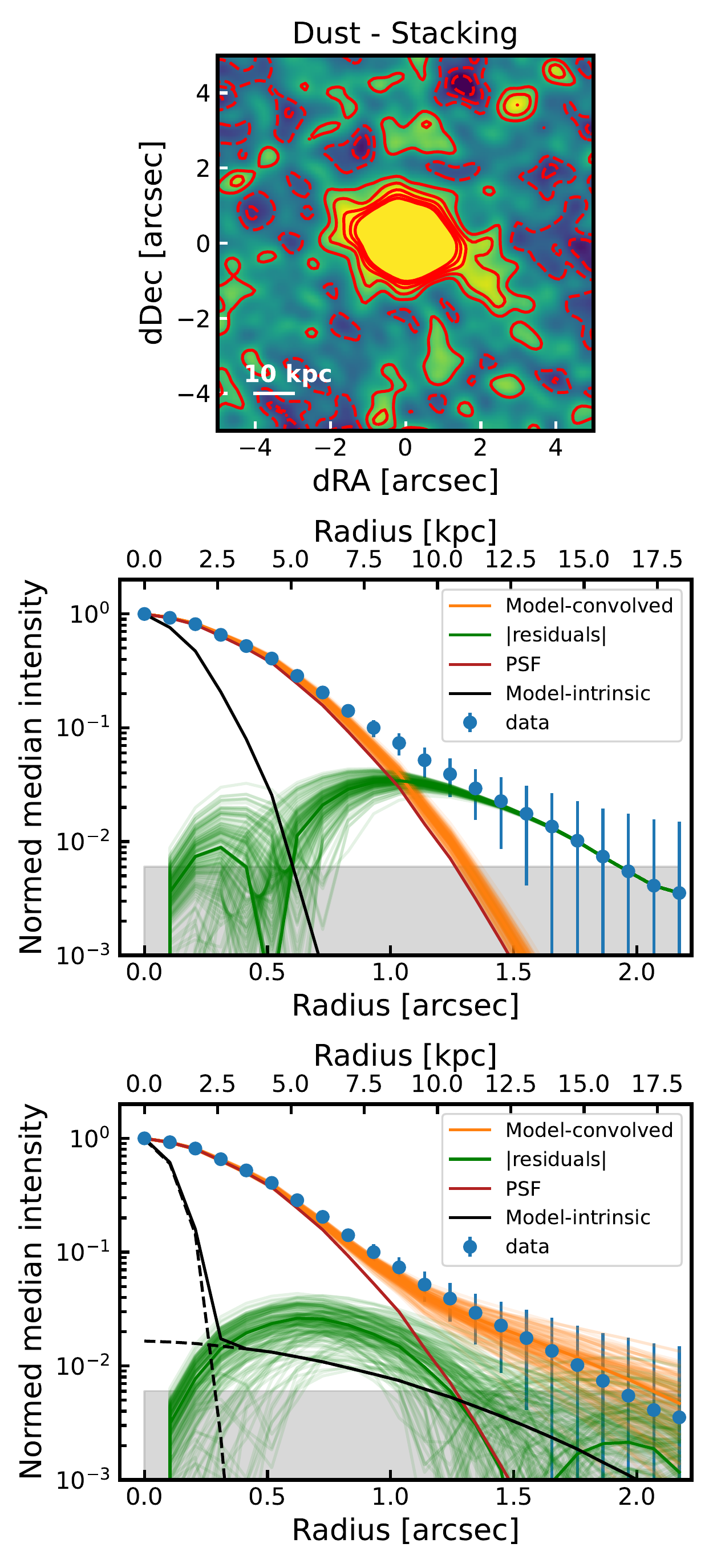}
   \caption{ Modelling of the aperture profiles of the dust continuum stacked emission. Top panel: The red solid contours show 1, 2, 3, 4, 5 $\sigma$ and the red dashed contours -1, -2, -3, -4, -5 $\sigma$. Middle and bottom panels: Modelling of the radial brightness profiles using a single resolved source model (middle panel) and two resolved sources model. The blue points show the extracted radial brightness profile and their uncertainties. The orange and green lines show 100 randomly drawn solutions from the MCMC chain for the fit and residuals, respectively. The black line shows the intrinsic model before the convolution.  The shaded region shows the 0.5$\times$ RMS of the moment-0 maps.
    }
   \label{fig:Cont_Aperture_modelling}
\end{figure}

We further investigate and model the image-plane stacking in Figures: \ref{fig:CO_Aperture_modelling}, \ref{fig:CI_Aperture_modelling}, \ref{fig:H2O_Aperture_modelling}, \ref{fig:Cont_Aperture_modelling}. In the top subplot of each Figure, we show moment-0 map of the stack from the velocity range which gives the most robust evidence for extended halo emission ($\pm 200$ km s$^{-1}$ for [\ion{C}{i}](2-1) and H$_2$O emission and $\pm 300$ km s$^{-1}$ for CO(7-6) emission). The extracted radial surface brightness profiles (second and third subplots) show emission on scales larger than the beam for CO(7-6), [\ion{C}{i}](2-1) and dust continuum.

\begin{figure*}
    \includegraphics[width=0.8\paperwidth]{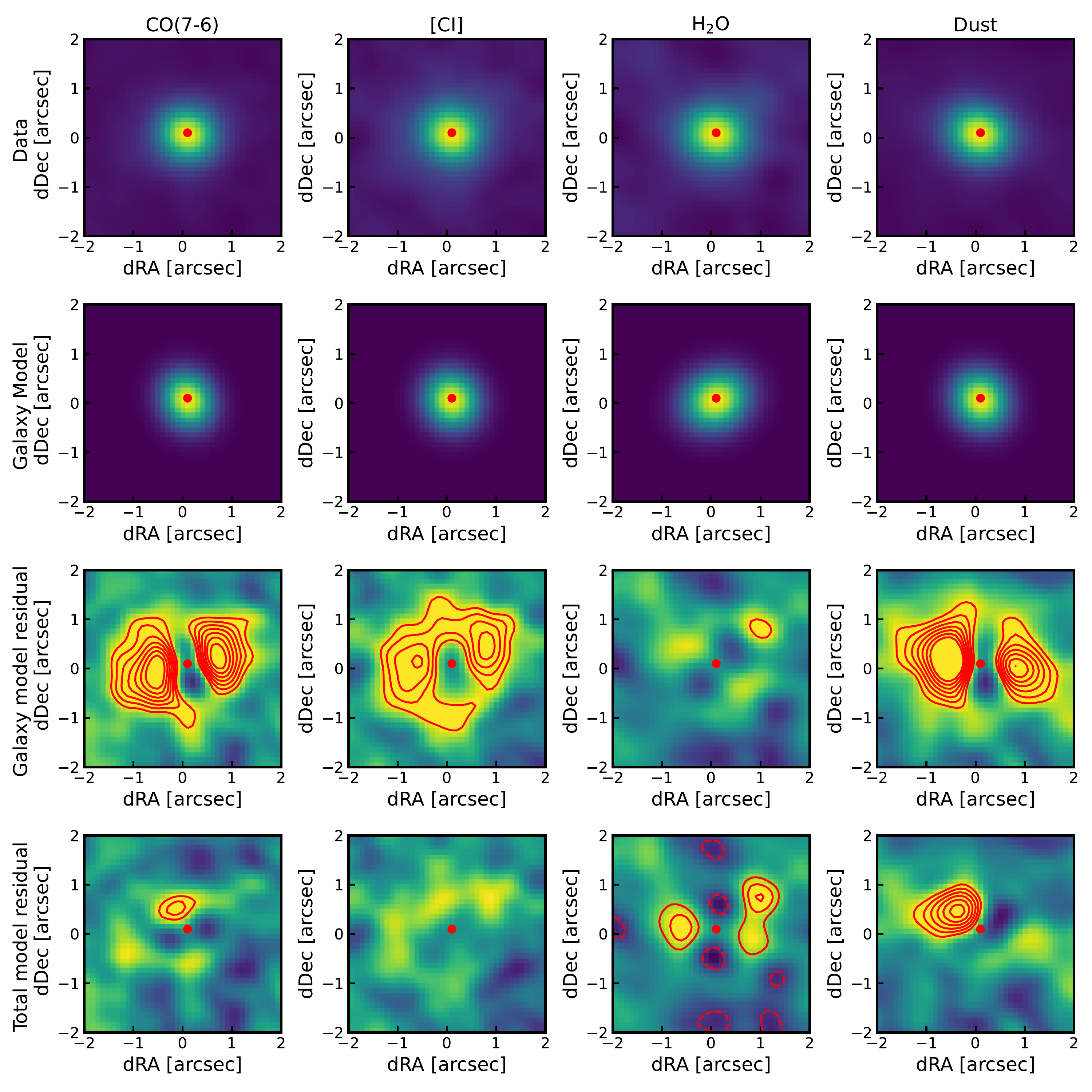}
   \caption{ Image visualisation of the aperture growth modelling. Each column represents different emission tracers. From left to right: CO(7-6), [\ion{C}{i}](2-1), H$_2$O an dust continuum. From top to bottom: First row: Data from the image-based stacking. The red dot indicates the centre of the image. Second row: Model image constructed from the best-fit to the radial brightness profiles convolved with the beam. The model is dominated by the galaxy component. Third row: Residual image after subtracting the galaxy component model. Fourth row: Residual image after subtracting the total model image (i.e. both galaxy and outer components). In each residual images, the solid contours show SNR levels of 2 and increasing by one and dashed contours show negative contours starting at -2. 
    }
   \label{fig:Stack_resid}
\end{figure*}

Using the methods described in \S~\ref{sec:RP_analyses}, we fitted the extracted radial brightness profiles with a single resolved galaxy component in the middle panels of Figures \ref{fig:CO_Aperture_modelling}, \ref{fig:CI_Aperture_modelling}, \ref{fig:H2O_Aperture_modelling}, \ref{fig:Cont_Aperture_modelling}. The residual of the fits (green solid line) shows significant emission on scale larger than one arcsecond for CO(7-6),  [\ion{C}{i}] and dust continuum stacked data. We show the image of the model galaxy component and the moment-0 residual in the second and third row of Figure \ref{fig:Stack_resid}, respectively, further showing the residual emission on scales of $>1$ arcsecond, that are not accounted for by the fitting a resolved galaxy component only.

We fitted the galaxy resolved galaxy component and an extended halo in the bottom panels of \ref{fig:CO_Aperture_modelling}, \ref{fig:CI_Aperture_modelling}, \ref{fig:H2O_Aperture_modelling}, \ref{fig:Cont_Aperture_modelling}. The BIC and reduced $\chi^{2}$ indicates that the two-component model is a better fit for the stacked CO(7-6), [\ion{C}{i}] and dust continuum data, while the single galaxy component is a better fit for the stacked H$_2$O data. We show the moment-0 residual from fitting the two-component model in the bottom row of Figure \ref{fig:Stack_resid}. We see small residual emission in the CO(7-6) and dust emissions, suggesting that the emission halo is not strictly symmetrical as we model. Overall, the residual images confirm the presence of large extended halos in the image-stacked data as shown in the aperture growth analysis. This is the first detection of a cold molecular gas emission in the CGM of ERQs at high redshift using emission lines.

The estimated deconvolved halo emission FWHM sizes are $13.5\pm0.66$, $12.6\pm1.24$ and $14.6\pm2.7$ kpc for CO(7-6), [\ion{C}{i}](2-1) and dust emission, respectively. These sizes are smaller than the previous detection of [\ion{C}{ii}] emission halos at $z>5$ of $\sim 22$ kpc \citep[][]{Fujimoto19,Fujimoto20} and CO(3-2) emission of $z\sim2.5$ AGN from the SUPER survey ($\sim27$\,kpc, \citealt{Jones22}). We do not detect any extended halo emission in H$_2$O, however, this is expected as H$_2$O emission is mostly tracing dense warm gas. We confirmed that the stacked emission is not dominated by the objects with individually detected extended emission (J1232+0912 and J2323-0100), by repeating the stacking procedure excluding these sources from the stacking. After removing the objects with individually detected halos from the stacks we measured the sizes of the halo component as $13.9\pm0.8$ and $12.2\pm1.78$ kpc for CO(7-6) and [\ion{C}{i}](2-1), respectively. Therefore, this detected extended emission is not from a single source but is present in all of the sources.

\begin{table*}
\caption{Radial surface brightness profile fitting results for stacked data and results of the modelling the \textit{uv}-visibilities of the stacked data. (1) Tracer in which we detect the extended emission; (2) Model fitted to the data (best fit model is in bold); (3) Size of the galaxy component from radial surface brightness profiles; (4) Size of the outer component from radial surface brightness profiles; (5) log$_{10}$ ratio between the peaks of the two-component from radial surface brightness profiles; (6) BIC; (7) Size of the galaxy component from \textit{uv}-visibilities fit; (8) Size of the outer component from \textit{uv}-visibilities fit; (9) log$_{10}$ ratio between the flux of the two-component from \textit{uv}-visibilities fit; (10) BIC
}
\begin{tabular}{lc|ccccc|cccc} 
    
\hline  
\hline  
     &       & \multicolumn{4}{c}{Image-based stacking} &  \multicolumn{4}{c}{uv-based stacking} \\
     (1) & (2) & (3) & (4) & (5) & (6) & (7) & (8) & (9) & (10) & (11) \\ 
Tracer & Model & FWHM(galaxy) & FWHM(halo) & log$_{10}$($\frac{\rm Peak_{\rm halo}}{\rm Peak_{\rm galaxy}}$) & BIC & $\chi^2$ & FWHM(galaxy) & FWHM(halo) & log$_{10}$($\frac{F_{\rm halo}}{F_{\rm galaxy}}$) & BIC \\ 
    &     &  [arcsec] & [arcsec] & & & & [arcsec] & [arcsec] & \\
\hline  

\multirow{2}{*}{CO(7-6)} & Single & $ 0.4 ^{+ 0.03 }_{- 0.03 }$  & - & - & 29.4 &26.0 & $ 0.21 ^{+ 0.01 }_{- 0.01 }$  & - & - & 153\\
                         & \textbf{Double} & $ 0.11 ^{+ 0.11 }_{- 0.07 }$  & $ 1.71 ^{+ 0.35 }_{- 0.24 }$  & $ -2.64 ^{+ 0.3 }_{- 0.21 }$ & 12.7 &3.4 & $ 0.19 ^{+ 0.01 }_{- 0.01 }$ & $ 2.3 ^{+ 0.33 }_{- 0.27 }$ & $-0.80\pm0.1$ &119 \\ [2pt]
\hline                          
\multirow{2}{*}{[CI]}    & Single & $ 0.6 ^{+ 0.05 }_{- 0.05 }$ & - & - & 16.5& 15.6 & $0.35 ^{+ 0.01 }_{- 0.01 }$ &  - & -  & 104\\
                         & \textbf{Double} &  $ 0.22 ^{+ 0.17 }_{- 0.15 }$ & $ 1.62 ^{+ 0.46 }_{- 0.31 }$  & $ -1.91 ^{+ 0.34 }_{- 0.42 }$ & 11.3 &2.9 & $ 0.31 ^{+ 0.02 }_{- 0.02 }$  & $ 1.66 ^{+ 0.4 }_{- 0.34 }$ & $-0.57\pm0.3$ & 82 \\ [2pt]
\hline 
\multirow{2}{*}{H$_2$O} & \textbf{Single} & $ 0.19 ^{+ 0.11 }_{- 0.12 }$ & - & - & 8.22 & 5.1 & $ 0.21 ^{+ 0.02 }_{- 0.02 }$ & - & - & 91\\
                        & Double &  $ 0.23 ^{+ 0.1 }_{- 0.13 }$ & $ 1.76 ^{+ 1.12 }_{- 0.57 }$  & $ -3.15 ^{+ 0.42 }_{- 0.26 }$ & 14.7& 5.4 & $ 0.2 ^{+ 0.02 }_{- 0.02 }$ &  $6.21 ^{+ 31.24 }_{- 5.42 }$ & $-1.14\pm0.70$ & 99\\ [2pt]
\hline 
\multirow{2}{*}{Dust} & Single & $ 0.41 ^{+ 0.03 }_{- 0.03 }$  & - & - &33.5 & 36.2 & $ 0.24 ^{+ 0.01 }_{- 0.01 }$  & - & - & 172\\
                      & \textbf{Double} & $0.15 ^{+ 0.11 }_{- 0.1 }$  & $ 1.99 ^{+ 0.57 }_{- 0.36 }$  & $ -2.62 ^{+ 0.29 }_{- 0.3 }$ & 19.7& 22.0 & $ 0.2 ^{+ 0.01 }_{- 0.01 }$  & $ 1.54 ^{+ 0.21 }_{- 0.21 }$  & $-0.64\pm0.28$ & 87\\ [2pt]
\hline  
\end{tabular} 
\label{Table:stack_res}
\end{table*}

\subsubsection{\textit{uv}-plane stacking results}\label{sec:uv_stack_res}

\begin{figure}
    \includegraphics[width=0.99\columnwidth]{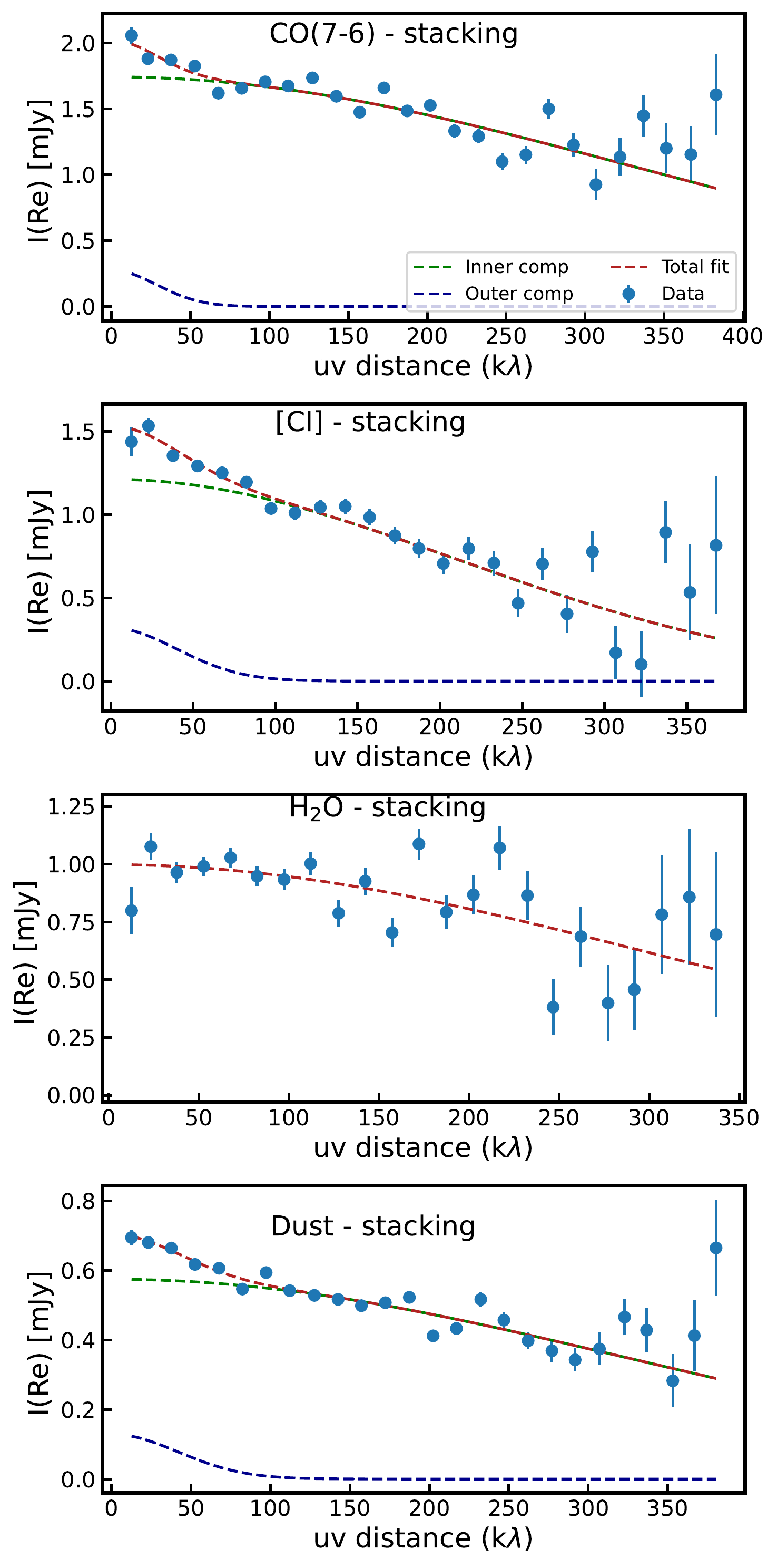}
   \caption{Results of the \textit{uv}-based stacking. Panels from top: CO(7-6), [\ion{C}{i}](2-1), H$_2$O and dust continuum. The stacked and binned \textit{uv}-visibilities are shown as blue points, while the best fit is shown as a red dashed line. When two separate components are required in the fit we show the individual components as a green and blue dashed line. The stacked \textit{uv}-visibilities show a presence of additional extended emission in CO(7-6), [\ion{C}{i}](2-1) and dust continuum data. 
    }
   \label{fig:uv_stacking}
\end{figure}

We stacked the targets in \textit{uv}-plane for all four emission tracers. We extracted the visibilities from the stacked measurement sets and we show these in Figure \ref{fig:uv_stacking}. Visually, all of these stacked \textit{uv}-visibilities show a Gaussian-like profile, indicating at least a single resolved source in the stacked data. However, in the CO(7-6), [\ion{C}{i}](2-1) and dust continuum \textit{uv}-stacked data, we also see a sudden upturn in flux density at low \textit{uv}-distances, indicating an additional large scale component.

In order to confirm the presence of extended emission, we fitted the collapsed stacked \textit{uv}-visibilities using three separate models: a single resolved source (half-Gaussian model), resolved source and a point source (half-Gaussian model + a constant), and two separate resolved sources (two half-Gaussian models). Based on the BIC and $\chi^{2}$, our modelling of the \textit{uv}-visibilities favoured two resolved components in CO(7-6), [\ion{C}{i}](2-1) and dust continuum stacks, while for the H$_2$O data, the BIC favours a single resolved model, in agreement with the image-based stacking. 

We measured the sizes of the extended (halo) component to be $18.4\pm2.4$, $13.3\pm3.2$ and $12.3\pm1.6$ kpc for CO(7-6), [\ion{C}{i}](2-1) and dust emission, respectively. These values agree with the results from the image-stacking within 1$\sigma$ error. Overall, the results of the \textit{uv}-stacking confirm our previous results of the existence of extended cold gas halos around these quasars. We summarise the \textit{uv}-based stacking results in Table \ref{Table:stack_res}.

\subsection{Estimating dust and molecular gas masses in the halos}\label{sec:mass_estimates}

%[There have been a ton of works recently using single-point SED models. Unfortunately, I'm picky about FIR SED fitting :)
%-In addition to saying that SMGs are optically thick, put in a quick calculation of optical depth, which is just a function of galaxy size and whatnot (the equation is in my 2020 paper)
%-Include CMB effects on dust temperature (will be small at z~2.5, but still).
%-Since TD and BetaIR are unknown (z~2.5 SFGs should have a very different dust T and the galactic BetaIR doesn't seem applicable), just use broad ranges of like TD~20-100, Beta~1-2.]

The \textit{uv}-stacking allowed us to reliably estimate the fluxes of the halo components and as a result, we can estimate the total dust and molecular gas mass. We derive the dust masses using a single modified black body (MBB) curve,  (e.g., \citealt{Jones20}) as:
\begin{equation}
S_{\rm obs}(\nu_{\rm obs})=\frac{(1+z)\pi R^2}{D_{\rm L}^2}B'(\nu,T_{\rm dust})\left( 1-e^{\frac{-M_{\rm dust}\kappa_o(\nu/\nu_o)^{\beta}}{\pi R^2}} \right)
\end{equation}
where $S_{\rm obs}(\nu_{\rm obs})$ is the observed band 6 flux, $R$ is the size of the galaxy or halo, $\kappa_o$ = 4 cm$^2$ g$^{-1}$ with $\nu_0$ = 1.2 THz \citep[see][]{Bianchi13} and $\beta$ = 1.8. As the source is at high redshift, it is necessary to also include the effect of the dust heating by the CMB. Removing this contribution results in a modified black body function ($B'(\nu,T_{\rm dust})$):
\begin{multline}
B'(\nu,T_{\rm dust})=B(\nu,T_{\rm dust}')-B(\nu,T_{\rm CMB})\\
= \frac{2h\nu^3}{c^2}\left[\frac{1}{e^{h\nu/k_BT_{\rm dust}'}-1}-\frac{1}{e^{h\nu/k_BT_{\rm CMB}}-1} \right]
\end{multline}
where $T_{\rm dust}$ is the true dust temperature, $T_{\rm dust}'$ is an effective dust temperature, $T_{\rm CMB}=(1+z) T_o$, with $T_{\rm o}=2.73$\,K is the CMB temperature at $z=0$. We assume dust temperature of 30 K and we discuss this value below and beta value of 1.8.%We assume a power law with dependence on frequency for k$_\nu$, k$_\nu$ = k$_{\nu0}$ ($\frac{\nu}{\nu_{{0}}}$)$^{\beta}$ cm$^{2}$ g$^{−1}$. We adopt k$_{\nu0}$ = 4 cm$^2$ g$^{-1}$ with $\nu_0$ = 1.2 THz \citep[see .][]{Bianchi13} and $\beta$ = 1.8, which is the Galactic value from the Planck data \citep[][]{Planck13}. 
%We assume that the dust is optically thin at the rest-frame wavelength (360 $\mu$m) of our observations as was shown in sub-mm galaxies (SMGs; $\lambda_{\rm rest}$ = 200 $\mu$m, \citealt{Conley11}, \citealt{Riechers13}; $\lambda_{\rm rest}$ = 100 $\mu$m, \citealt{Simpson17}).

For estimating the cold gas mass in the host galaxies and in the halo we use the [\ion{C}{i}](2-1) emission line. The CO(7-6) traces more excited gas (150 K ) and the conversion to CO(1-0) necessary to estimate cold gas mass is very uncertain, even in star-forming galaxies, let alone in uncertain conditions of these extended halos. To calculate the cold gas mass ($M$(H$_2$)) from [\ion{C}{i}](2-1), we use the method described in \citet{Bothwell17}. We calculate $M$(H$_2$) as:
\begin{equation}
  \begin{aligned}
    M(H_{2}) = 1375.8 \times D^2_L (1+z)^{-1} \left(\frac{X_{[\ion{C}{i}]}}{10^{-5}}\right)^{-1} \left(\frac{A_{10}}{10^{-7} s^{-1}}\right)^{-1}\\
    \times Q_{10}^{-1} I_{\rm [\ion{C}{i}]_{(1-0)}}
 \end{aligned}
\end{equation}
where $X_{\rm[CI_{1-0}]}$ is the [\ion{C}{i}]/H$_2$ abundance ratio, we adopt a literature-standard [\ion{C}{i}]/H2 abundance ratio of $3\times 10^{-5}$ and the Einstein A coefficient ($A_{10}$) of $7.93 \times 10^{-8}$s$^{-1}$ with excitation  factor ($Q_{10}$) of 0.6. The value of $Q_{10}$ is dependent on the specific conditions within the gas \citep[][]{Papadopoulos04}. As we observed the [\ion{C}{i}](2-1) line rather than the [\ion{C}{i}](1-0) and hence we need to convert these, using the [\ion{C}{i}](1-0)/[\ion{C}{i}](2-1) conversion factor of 3 \citep[][]{Jiao17}

Using the method and assumptions above, we estimated average dust mass of $10^{8.3 \pm 0.16}$ and $10^{7.6 \pm 0.12}$ M$_{\odot}$ for the galaxy and halo components, respectively, and average molecular gas masses of $10^{10.8\pm 0.14}$ and $10^{10.2\pm 0.16}$ M$_{\odot}$ for the galaxy and halo components, respectively. We note that the quoted uncertainties are estimated from the random flux uncertainties. We discuss the systematic uncertainties below. The average host galaxy masses are in agreement with molecular gas and dust masses measured in high-z quasars \citep[][]{Bischetti21, Decarli22}. These halo molecular gas masses indicate a massive cold gas reservoir around these luminous quasars. Previous studies estimating these molecular gas halos around star-forming galaxies estimated $10^{11.3-11.77}$M$\odot$ \citep[with a range of CO(4-3)/CO(1-0) and $\alpha_{\rm CO}=10$;][]{Ginolfi17} and $1.3\times 10^{11}$M$\odot$ \citep[using CO(4-3)/CO(1-0) = 0.45--1 and $\alpha_{\rm CO}$=3.6;][]{Li21}. 

Here, we discuss the uncertainties in estimating the dust and molecular gas masses, which in both cases, are dominated by systematic rather than random uncertainties. The primary source of uncertainties in estimating the dust masses is the assumed dust temperature of 30 K, as this is a typical value for $z\sim$ 2.5 star-forming galaxies \citep[30-40 K; ][]{Schreiber18,Liang19, Reuter20}.  However, the dust in the halo is located away from any of the heating sources such as star-formation and quasar, it can be significantly cooler than dust in the galaxy. Recalculating the dust mass in the halos for a temperature of 20 K (lowest measured temperature for SF galaxy at $z\sim$2.4; \citealt{Reuter20}) yields $10^{8.3}$ M$_{\odot}$, a factor of 5 higher than the original estimate for the dust mass in the halo. 

There have been numerous studies focusing on the [\ion{C}{i}] abundance and its effect on estimates of the molecular gas mass. Although [\ion{C}{i}]  is certainly more stable and more reliable than the CO for determining the molecular gas mass, the [\ion{C}{i}]-to-H$_2$ conversion factor is also associated with uncertainties. Both \citet{Offner14} and \citet{Glover16} show results from post-processing of hydrodynamical simulations of star-forming clouds, claiming that the [\ion{C}{i}] abundance varies as a function of interstellar radiation field (ISRF), metallicity and H$_2$ column density. At high A$_{\rm v}$, the [\ion{C}{i}] abundance is raised by the presence of cosmic rays, while at low A$_{\rm v}$,  increasing the ISFR by a factor of 100--1000 can increase the [\ion{C}{i}] abundance by 30--50\%, a similar effect to $\alpha_{\rm CO}$. Furthermore, \citet{Glover16} have showed evidence for evolving [\ion{C}{i}] abundance as a function of metallicity given as: $\sim \rm Z^{-1}$. The metallicity of CGM material can vastly vary. Indeed, \citet{Pointon19} found that the metallicity of cooler CGM ($10^{4}$ K) can vary by a factor 10--100, and there are no measurements of the metallicity of $<100$\,K CGM. However, if we consider a range of metallicity values for our molecular gas measurement, the average molecular gas mass in these cold gas halos can be up to $10^{12.1}$ M$\odot$ (for log$_{10}$ OH = -1.5). However, given that the origin of this cold is most likely AGN or star formation driven outflows, the metallicity of the gas is going to be closer to that of a galaxy.

An alternative approach to calculating molecular gas masses is to use the dust mass as a proxy tracer of the molecular gas \citep[e.g.][]{Leroy11, Magdis11, Magnelli12, Scoville14, Genzel15}. Therefore, assuming a gas-to-dust ratio ($\delta$GDR), molecular gas masses can be estimated as M($ H_{2}$ = $\delta$GDR M$_{\rm dust}$. As mentioned above, the lack of metallicity estimates of the cold CGM, we adopt a fixed value of $\delta$GDR = 100-1000 \citep[for solar metallicity star-forming galaxies ][]{Sandstrom13, Remy-Ruyer14}. We estimate molecular gas masses from dust measurement of $10^{9.6-10.6}$ M$_{\odot}$, within the estimated values using the [\ion{C}{i}](2-1) emission. 

Overall, we estimated a range of average cold gas masses inside these halos to be in the range of $10^{10.2}-10^{12.1}$ M$\odot$. This indicates that these quasars have significant gas reservoirs surrounding their host galaxies. In the next section, we discuss the origin of these cold gas halos.

\subsection{Origins of the halo emission}\label{sec:origins}

In this section we discuss the origin of these extended cold gas halos. The potential origins of these halos are: 1) current merger events; 2) companions; 3) cold gas halos from AGN or SF driven outflows. To distinguish between these scenarios we investigated the emission line kinematics and HST archival i-band and 1.4 $\mu$m imaging of our targets. Given the double peak nature of the emission lines (see Figures \ref{fig:CO_data}, \ref{fig:CI_data} \& \ref{fig:H2O_data}) and smooth velocity gradient of moment-1 maps, we see no evidence of disturbed kinematics suggesting a recent merger. 

\begin{figure}
    \includegraphics[width=0.99\columnwidth]{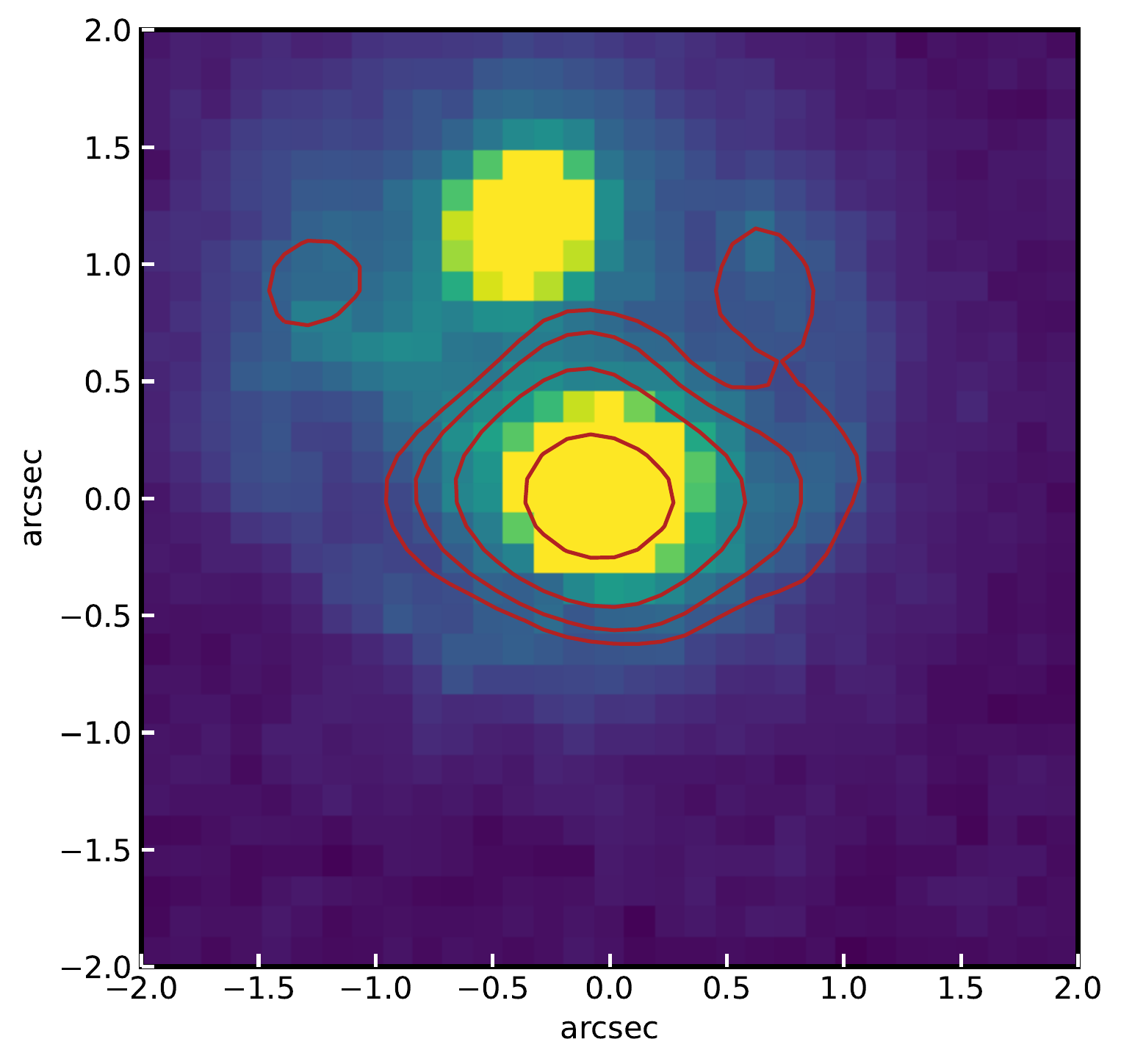}
   \caption{ HST $1.4\mu$m image of the J2323-0100 with ALMA  [\ion{C}{i}] map overlayed as red solid (2,3, 5 and 10 $\sigma$ levels) contours. The image is centred on the location of the quasar. The HST imaging show clear signs of a merger, suggesting that the detection of large scale cold gas emission in this particular object is caused by the merger, rather than a cold gas halo (see Figure \ref{fig:232326_CI_Aperture_modelling}).
    }
   \label{fig:HST_merger}
\end{figure}

Overall, six targets have deep HST i-band and $1.4 \mu$m imaging, with five of six targets showing no sign of additional objects within 1.5 arcsecond. However, in Figure \ref{fig:HST_merger} we present HST 1.4$\mu$m map ($\lambda_{\rm rest=}$400 \AA) imaging of J2323-0100 with ALMA [\ion{C}{i}] contours overlaid in red. The presence of an additional bright object 1.2 arcsecond away from the main quasar with additional faint emission possibly around the bright point source may indicate a galaxy merger with tidal stellar streams between them \citep[see also][ for merger discussion]{Vayner21}. As a result, we cannot confirm that the individually detected large-scale cold gas emission in J2323-0100 is a cold gas halo. For the rest of the targets, we see no evidence of mergers in either the emission line kinematics or HST imaging. Furthermore, given that we do not see any evidence for companion galaxies in the HST imaging and we do not detect the extended emission in H$_2$O emission, we conclude that our extended emission are indeed cold gas halos, rather than contamination from mergers or companions.

We investigated the ionised outflow velocities of the two ERQs with individually detected large scale cold gas halos. \citet{Perrotta19} have observed the ionised outflow velocities in [\ion{O}{iii}] of 4800 and 3000 km s$^{-1}$. Although these are extreme velocities, they are in a common range for high luminosity ERQs such as our sample \citep[][]{Perrotta19, Vayner21}. 

It is now important to discuss the origin of the cold gas halo. Previous studies \citep[][]{Fujimoto19, Fujimoto20, Ginolfi20}) showed evidence of cold gas halos detected in  [\ion{C}{ii}] emission in star-forming galaxies, suggesting that the origin of the halos can be starburst driven winds. A possibility, cosmological simulations showed that bright galaxies at high redshift were likely encased in diffuse filaments of gas (e.g., \citealt{Pallotini17}, \citealt{Kohandel19}). However, the presence of dust and [\ion{C}{i}] emission in these halos indicates significant metal enrichment of these halos, pointing towards its origin in the galaxy.

Assuming molecular gas outflow velocities of 700--2000 km s$^{-1}$ \citep[][]{Bischetti19, Stanley19}, we calculated the travel times of 7-35 Myr for the gas to reach such distances from the host galaxy. As quasar and AGN can switch can vary drastically on scales of 1-10 Myr \citep[e.g.][]{Hickox14, Schawinski15, King15, McAlpine17}, it is less likely that the current quasar episode is responsible for the creation of these cold gas halos, but rather than they are relics of previous AGN or starbursts episodes. However, assuming the velocity gas similar to that of the detected ionised outflows in these objects \citep[$\sim$ 3000 km s$^{-1}$;][]{Vayner21}, the time taken for the gas to reach 13 kpc would only be 4.2 Myr. However, this would require this ionised gas to cool down to 40 K, traced by [\ion{C}{i}](2-1) emission, in the same time.  

As the [\ion{C}{i}](2-1) line is tracing the cold phase of the gas (T$\sim$30 K) and the CO(7-6) is tracing excited dense gas (T$\sim$150 K), we speculate that the cold gas halos contain a large amount of both cold and warm excited gas, as both [\ion{C}{i}] and CO(7-6) have similar halo extension. Finally, these metal-enriched cold molecular gas halos can be either created by AGN or star-formation driven outflows \citep[][]{Maiolino12, Cicone15, Fiore17, Spilker18}. The most likely scenario is that these halos were created by a combination of past AGN and starburst activity. Overall, this is the first detection of cold molecular gas halos traced in  [\ion{C}{i}], CO(7-6) and dust continuum emission in high redshift galaxies. Our result confirms the predictions of the cosmological simulations that the baryon cycle the enriched gas exchanges with the CGM are at work \citep[see e.g., ][]{Hopkins14, Somerville15, Hayward17}. 

\section{Conclusions}

We present the results of ALMA Band 6 observations of 15 extremely red quasars. We detect the 13, 11, 10 and 13 objects in CO(7-6), [\ion{C}{i}](2-1), H$_2$O and continuum emission, with SNR ranging from 5 to 64 $\sigma$. We constructed a radial brightness profile for both individual sources and for stacked data (in both image and \textit{uv}-plane) to search for extended emission around these objects.  Based on our analyses we find:

\begin{enumerate}
    \item We measured the sizes of the four emission tracers to be in the range of 3.1--8.6 kpc (median of 4.0 kpc) for CO(7-6), 2.9--11.2 kpc (median of 5.3 kpc) for [\ion{C}{i}](2-1), 2.7--6.0 kpc (median of 4.3 kpc) for H$_2$O and 2.0--8.0 kpc (median of 5.1 kpc) for continuum emission. These value are consistent with those found in the literature for sub-mm galaxies and other AGN host galaxies (see Figure \ref{fig:Size_comp}). 
    
    \item Modelling the observed radial surface brightness profiles, we found extended emission in two objects in either CO(7-6) or [\ion{C}{i}](2-1) emission (see Figures \ref{fig:123241_CO_Aperture_modelling} \& \ref{fig:232326_CI_Aperture_modelling}). We measured the FWHM sizes of these extended halos to be $21.6^{+5.0}_{-4.0}$ kpc for the CO(7-6) halo in J1232+0912 and $19.8^{+7.2}_{-5.0}$ kpc for the [\ion{C}{i}](2-1) halo in J2323+0100. 
    
    \item We stacked our sample in CO(7-6), [\ion{C}{i}](2-1), H$_2$O and dust continuum emission in the image plane and extracted the the radial surface brightness profiles from the  moment-0 maps of the stacked emission cubes (see Figures \ref{fig:CO_Aperture_modelling}, \ref{fig:CI_Aperture_modelling}, \ref{fig:H2O_Aperture_modelling} \& \ref{fig:Cont_Aperture_modelling}). Modelling the profiles showed evidence of large scales cold gas halos in CO(7-6), [\ion{C}{i}](2-1) and dust continuum with size of $13.5\pm0.66$, $12.6\pm1.24$ and $14.6\pm2.7$ kpc for CO(7-6), [\ion{C}{i}](2-1) and dust emission, respectively. Investigating the residual stacked images after subtracting a central galaxy source (see Figure \ref{fig:Stack_resid}) confirms the radial surface brightness profile modelling. 
    
    \item Stacking our data in the \textit{uv}-plane across the four emission tracers and extracting the \textit{uv}-visibilities confirms the result of the image stacking extended cold gas halos around these quasar host galaxies (see Figure \ref{fig:uv_stacking}). We measure the sizes of the halo component to be $18.4\pm2.4$, $13.3\pm3.2$ and $12.3\pm1.6$ kpc for CO(7-6), [\ion{C}{i}](2-1) and dust emission, respectively. These cold gas halo sizes agree within 1$\sigma$ with the sizes measured in the image-plane.
    
    \item Using the measured fluxes of the dust continuum and the [\ion{C}{i}](2-1) emission line from the \textit{uv}-plane stacking we derived the average dust and molecular gas mass inside the halo of $10^{7.6}$ and $10^{10.2}$ M$_{\odot}$, respectively. These dust and molecular gas masses indicate substantial dust and gas reservoirs around these quasar host galaxies and evidence of enrichment of CGM from the past AGN or starbursts activity. 
    
\end{enumerate}

Overall, our analysis of this deep ALMA band 6 data shows evidence for a central host galaxy source surrounded by an extended cold gas and dust halo. Assuming typical molecular and ionised gas outflow velocities implies long travel times for the gas to reach such distances (5-30 Myr), and hence suggesting these halos are relics of past AGN or star-formation activity. 

\section*{Data Availability}
The datasets were derived from sources in the public domain: ALMA data from \url{https://almascience.nrao.edu/aq/
?result_view=observation}. The images, spectra and stacked data in this article will be shared on reasonable request to the corresponding author.

\section*{Acknowledgements}

J.S. and R.M. acknowledge ERC Advanced Grant 695671 “QUENCH” and support by the Science and Technology Facilities Council (STFC).
G.C.J. acknowledges funding from ERC Advanced Grants 789056 ``FirstGalaxies’’ under the European Union’s Horizon 2020 research and innovation programme.  
S.C is supported by European Union’s
HE ERC Starting Grant No. 101040227 - WINGS. 
This paper makes use of ALMA data: 2017.1.00478.S.
ALMA is a partnership of ESO (representing its member states), NSF (USA) and NINS (Japan), together with NRC (Canada) and NSC and ASIAA (Taiwan), in cooperation with the Republic of Chile. The Joint ALMA Observatory is operated by ESO, AUI/NRAO and NAOJ.


\bibliographystyle{mnras}
\bibliography{mybib} % if your bibtex file is called mybib.tex

\begin{thebibliography}{}
\makeatletter
\relax
\def\mn@urlcharsother{\let\do\@makeother \do\$\do\&\do\#\do\^\do\_\do\%\do\~}
\def\mn@doi{\begingroup\mn@urlcharsother \@ifnextchar [ {\mn@doi@}
  {\mn@doi@[]}}
\def\mn@doi@[#1]#2{\def\@tempa{#1}\ifx\@tempa\@empty \href
  {http://dx.doi.org/#2} {doi:#2}\else \href {http://dx.doi.org/#2} {#1}\fi
  \endgroup}
\def\mn@eprint#1#2{\mn@eprint@#1:#2::\@nil}
\def\mn@eprint@arXiv#1{\href {http://arxiv.org/abs/#1} {{\tt arXiv:#1}}}
\def\mn@eprint@dblp#1{\href {http://dblp.uni-trier.de/rec/bibtex/#1.xml}
  {dblp:#1}}
\def\mn@eprint@#1:#2:#3:#4\@nil{\def\@tempa {#1}\def\@tempb {#2}\def\@tempc
  {#3}\ifx \@tempc \@empty \let \@tempc \@tempb \let \@tempb \@tempa \fi \ifx
  \@tempb \@empty \def\@tempb {arXiv}\fi \@ifundefined
  {mn@eprint@\@tempb}{\@tempb:\@tempc}{\expandafter \expandafter \csname
  mn@eprint@\@tempb\endcsname \expandafter{\@tempc}}}

\bibitem[\protect\citeauthoryear{{Alaghband-Zadeh}, {Banerji}, {Hewett}  \&
  {McMahon}}{{Alaghband-Zadeh} et~al.}{2016}]{AlaghbandZadeh16}
{Alaghband-Zadeh} S.,  {Banerji} M.,  {Hewett} P.~C.,   {McMahon} R.~G.,  2016,
  \mn@doi [\mnras] {10.1093/mnras/stw682}, \href
  {https://ui.adsabs.harvard.edu/abs/2016MNRAS.459..999A} {459, 999}

\bibitem[\protect\citeauthoryear{{Alexander} \& {Hickox}}{{Alexander} \&
  {Hickox}}{2012}]{Alexander12}
{Alexander} D.~M.,  {Hickox} R.~C.,  2012, \mn@doi [\nar]
  {10.1016/j.newar.2011.11.003}, \href
  {http://adsabs.harvard.edu/abs/2012NewAR..56...93A} {56, 93}

\bibitem[\protect\citeauthoryear{{Arrigoni Battaia} et~al.,}{{Arrigoni Battaia}
  et~al.}{2018}]{ArrigoniBattaia18}
{Arrigoni Battaia} F.,  et~al., 2018, arXiv e-prints, \href
  {https://ui.adsabs.harvard.edu/abs/2018arXiv181010140A} {p. arXiv:1810.10140}

\bibitem[\protect\citeauthoryear{{Beckmann} et~al.,}{{Beckmann}
  et~al.}{2017}]{Beckmann17}
{Beckmann} R.~S.,  et~al., 2017, preprint, \href
  {http://adsabs.harvard.edu/abs/2017arXiv170107838B} {} (\mn@eprint {arXiv}
  {1701.07838})

\bibitem[\protect\citeauthoryear{{Bianchi}}{{Bianchi}}{2013}]{Bianchi13}
{Bianchi} S.,  2013, \mn@doi [\aap] {10.1051/0004-6361/201220866}, \href
  {https://ui.adsabs.harvard.edu/abs/2013A&A...552A..89B} {552, A89}

\bibitem[\protect\citeauthoryear{{Bisbas}, {Schruba}  \& {van
  Dishoeck}}{{Bisbas} et~al.}{2019}]{Bisbas19}
{Bisbas} T.~G.,  {Schruba} A.,   {van Dishoeck} E.~F.,  2019, \mn@doi [\mnras]
  {10.1093/mnras/stz405}, \href
  {https://ui.adsabs.harvard.edu/abs/2019MNRAS.485.3097B} {485, 3097}

\bibitem[\protect\citeauthoryear{{Bischetti}, {Maiolino}, {Carniani}, {Fiore},
  {Piconcelli}  \& {Fluetsch}}{{Bischetti} et~al.}{2019}]{Bischetti19}
{Bischetti} M.,  {Maiolino} R.,  {Carniani} S.,  {Fiore} F.,  {Piconcelli} E.,
   {Fluetsch} A.,  2019, \mn@doi [\aap] {10.1051/0004-6361/201833557}, \href
  {https://ui.adsabs.harvard.edu/abs/2019A&A...630A..59B} {630, A59}

\bibitem[\protect\citeauthoryear{{Bischetti} et~al.,}{{Bischetti}
  et~al.}{2021}]{Bischetti21}
{Bischetti} M.,  et~al., 2021, \mn@doi [\aap] {10.1051/0004-6361/202039057},
  \href {https://ui.adsabs.harvard.edu/abs/2021A&A...645A..33B} {645, A33}

\bibitem[\protect\citeauthoryear{{Blanton} et~al.,}{{Blanton}
  et~al.}{2017}]{Blanton+17}
{Blanton} M.~R.,  et~al., 2017, \mn@doi [\aj] {10.3847/1538-3881/aa7567}, \href
  {https://ui.adsabs.harvard.edu/abs/2017AJ....154...28B} {154, 28}

\bibitem[\protect\citeauthoryear{{Bolatto}, {Wolfire}  \& {Leroy}}{{Bolatto}
  et~al.}{2013}]{Bolatto13}
{Bolatto} A.~D.,  {Wolfire} M.,   {Leroy} A.~K.,  2013, \mn@doi [\araa]
  {10.1146/annurev-astro-082812-140944}, \href
  {https://ui.adsabs.harvard.edu/abs/2013ARA&A..51..207B} {51, 207}

\bibitem[\protect\citeauthoryear{{Bothwell} et~al.,}{{Bothwell}
  et~al.}{2013}]{Bothwell13}
{Bothwell} M.~S.,  et~al., 2013, \mn@doi [\mnras] {10.1093/mnras/sts562}, \href
  {https://ui.adsabs.harvard.edu/abs/2013MNRAS.429.3047B} {429, 3047}

\bibitem[\protect\citeauthoryear{{Bothwell} et~al.,}{{Bothwell}
  et~al.}{2017}]{Bothwell17}
{Bothwell} M.~S.,  et~al., 2017, \mn@doi [\mnras] {10.1093/mnras/stw3270},
  \href {https://ui.adsabs.harvard.edu/abs/2017MNRAS.466.2825B} {466, 2825}

\bibitem[\protect\citeauthoryear{{Calistro Rivera} et~al.,}{{Calistro Rivera}
  et~al.}{2018}]{Callistro-Rivera18}
{Calistro Rivera} G.,  et~al., 2018, \mn@doi [\apj] {10.3847/1538-4357/aacffa},
  \href {https://ui.adsabs.harvard.edu/abs/2018ApJ...863...56C} {863, 56}

\bibitem[\protect\citeauthoryear{{Chabrier}}{{Chabrier}}{2003}]{Chabrier03}
{Chabrier} G.,  2003, \mn@doi [\pasp] {10.1086/376392}, \href
  {http://adsabs.harvard.edu/abs/2003PASP..115..763C} {115, 763}

\bibitem[\protect\citeauthoryear{{Chen} et~al.,}{{Chen} et~al.}{2017}]{Chen17}
{Chen} C.-C.,  et~al., 2017, \mn@doi [\apj] {10.3847/1538-4357/aa863a}, \href
  {http://adsabs.harvard.edu/abs/2017ApJ...846..108C} {846, 108}

\bibitem[\protect\citeauthoryear{{Chen} et~al.,}{{Chen} et~al.}{2020}]{Chen20}
{Chen} C.-C.,  et~al., 2020, \mn@doi [\aap] {10.1051/0004-6361/201936286},
  \href {https://ui.adsabs.harvard.edu/abs/2020A&A...635A.119C} {635, A119}

\bibitem[\protect\citeauthoryear{{Choi}, {Somerville}, {Ostriker}, {Naab}  \&
  {Hirschmann}}{{Choi} et~al.}{2018}]{Choi18}
{Choi} E.,  {Somerville} R.~S.,  {Ostriker} J.~P.,  {Naab} T.,   {Hirschmann}
  M.,  2018, \mn@doi [\apj] {10.3847/1538-4357/aae076}, \href
  {https://ui.adsabs.harvard.edu/abs/2018ApJ...866...91C} {866, 91}

\bibitem[\protect\citeauthoryear{{Cicone} et~al.,}{{Cicone}
  et~al.}{2015}]{Cicone15}
{Cicone} C.,  et~al., 2015, \mn@doi [\aap] {10.1051/0004-6361/201424980}, \href
  {https://ui.adsabs.harvard.edu/abs/2015A&A...574A..14C} {574, A14}

\bibitem[\protect\citeauthoryear{{Cicone} et~al.,}{{Cicone}
  et~al.}{2021}]{Cicone21}
{Cicone} C.,  et~al., 2021, \mn@doi [\aap] {10.1051/0004-6361/202141611}, \href
  {https://ui.adsabs.harvard.edu/abs/2021A&A...654L...8C} {654, L8}

\bibitem[\protect\citeauthoryear{{Cochrane} et~al.,}{{Cochrane}
  et~al.}{2019}]{Cochrane19}
{Cochrane} R.~K.,  et~al., 2019, \mn@doi [\mnras] {10.1093/mnras/stz1736},
  \href {https://ui.adsabs.harvard.edu/abs/2019MNRAS.488.1779C} {488, 1779}

\bibitem[\protect\citeauthoryear{{Concas} \& {Popesso}}{{Concas} \&
  {Popesso}}{2019}]{Concas19}
{Concas} A.,  {Popesso} P.,  2019, \mn@doi [\mnras] {10.1093/mnrasl/slz065},
  \href {https://ui.adsabs.harvard.edu/abs/2019MNRAS.486L..91C} {486, L91}

\bibitem[\protect\citeauthoryear{{Crain} et~al.,}{{Crain}
  et~al.}{2015}]{Crain15}
{Crain} R.~A.,  et~al., 2015, \mn@doi [\mnras] {10.1093/mnras/stv725}, \href
  {http://adsabs.harvard.edu/abs/2015MNRAS.450.1937C} {450, 1937}

\bibitem[\protect\citeauthoryear{{De Breuck}, {Lundgren}, {Emonts}, {Kolwa},
  {Dannerbauer}  \& {Lehnert}}{{De Breuck} et~al.}{2022}]{DeBreuck+22}
{De Breuck} C.,  {Lundgren} A.,  {Emonts} B.,  {Kolwa} S.,  {Dannerbauer} H.,
  {Lehnert} M.,  2022, \mn@doi [\aap] {10.1051/0004-6361/202141853}, \href
  {https://ui.adsabs.harvard.edu/abs/2022A&A...658L...2D} {658, L2}

\bibitem[\protect\citeauthoryear{{Decarli} et~al.,}{{Decarli}
  et~al.}{2022}]{Decarli22}
{Decarli} R.,  et~al., 2022, \mn@doi [\aap] {10.1051/0004-6361/202142871},
  \href {https://ui.adsabs.harvard.edu/abs/2022A&A...662A..60D} {662, A60}

\bibitem[\protect\citeauthoryear{{Delhaize}, {Meyer}, {Staveley-Smith}  \&
  {Boyle}}{{Delhaize} et~al.}{2013}]{Delhaize13}
{Delhaize} J.,  {Meyer} M.~J.,  {Staveley-Smith} L.,   {Boyle} B.~J.,  2013,
  \mn@doi [\mnras] {10.1093/mnras/stt810}, \href
  {https://ui.adsabs.harvard.edu/abs/2013MNRAS.433.1398D} {433, 1398}

\bibitem[\protect\citeauthoryear{{Di Matteo}, {Springel}  \& {Hernquist}}{{Di
  Matteo} et~al.}{2005}]{DiMatteo05}
{Di Matteo} T.,  {Springel} V.,   {Hernquist} L.,  2005, \mn@doi [\nat]
  {10.1038/nature03335}, \href
  {http://adsabs.harvard.edu/abs/2005Natur.433..604D} {433, 604}

\bibitem[\protect\citeauthoryear{{Drake} et~al.,}{{Drake}
  et~al.}{2020}]{drak20}
{Drake} A.~B.,  et~al., 2020, \mn@doi [\apj] {10.3847/1538-4357/aba832}, \href
  {https://ui.adsabs.harvard.edu/abs/2020ApJ...902...37D} {902, 37}

\bibitem[\protect\citeauthoryear{{Dubois}, {Pichon}, {Devriendt}, {Silk},
  {Haehnelt}, {Kimm}  \& {Slyz}}{{Dubois} et~al.}{2013a}]{Dubois13}
{Dubois} Y.,  {Pichon} C.,  {Devriendt} J.,  {Silk} J.,  {Haehnelt} M.,  {Kimm}
  T.,   {Slyz} A.,  2013a, \mn@doi [\mnras] {10.1093/mnras/sts224}, \href
  {https://ui.adsabs.harvard.edu/abs/2013MNRAS.428.2885D} {428, 2885}

\bibitem[\protect\citeauthoryear{{Dubois}, {Gavazzi}, {Peirani}  \&
  {Silk}}{{Dubois} et~al.}{2013b}]{Dubois13b}
{Dubois} Y.,  {Gavazzi} R.,  {Peirani} S.,   {Silk} J.,  2013b, \mn@doi
  [\mnras] {10.1093/mnras/stt997}, \href
  {https://ui.adsabs.harvard.edu/abs/2013MNRAS.433.3297D} {433, 3297}

\bibitem[\protect\citeauthoryear{{Emonts} et~al.,}{{Emonts}
  et~al.}{2016}]{Emonts+16}
{Emonts} B.~H.~C.,  et~al., 2016, \mn@doi [Science] {10.1126/science.aag0512},
  \href {https://ui.adsabs.harvard.edu/abs/2016Sci...354.1128E} {354, 1128}

\bibitem[\protect\citeauthoryear{{Emonts} et~al.,}{{Emonts}
  et~al.}{2018}]{Emonts+18}
{Emonts} B.~H.~C.,  et~al., 2018, \mn@doi [\mnras] {10.1093/mnrasl/sly034},
  \href {https://ui.adsabs.harvard.edu/abs/2018MNRAS.477L..60E} {477, L60}

\bibitem[\protect\citeauthoryear{{Fiore} et~al.,}{{Fiore}
  et~al.}{2017}]{Fiore17}
{Fiore} F.,  et~al., 2017, \mn@doi [\aap] {10.1051/0004-6361/201629478}, \href
  {https://ui.adsabs.harvard.edu/abs/2017A&A...601A.143F} {601, A143}

\bibitem[\protect\citeauthoryear{{F{\"o}rster Schreiber} et~al.,}{{F{\"o}rster
  Schreiber} et~al.}{2018}]{ForsterSch18a}
{F{\"o}rster Schreiber} N.~M.,  et~al., 2018, \mn@doi [\apjs]
  {10.3847/1538-4365/aadd49}, \href
  {http://adsabs.harvard.edu/abs/2018ApJS..238...21F} {238, 21}

\bibitem[\protect\citeauthoryear{{Fujimoto} et~al.,}{{Fujimoto}
  et~al.}{2018}]{Fujimoto18}
{Fujimoto} S.,  et~al., 2018, \mn@doi [\apj] {10.3847/1538-4357/aac6c4}, \href
  {http://adsabs.harvard.edu/abs/2018ApJ...861....7F} {861, 7}

\bibitem[\protect\citeauthoryear{{Fujimoto} et~al.,}{{Fujimoto}
  et~al.}{2019}]{Fujimoto19}
{Fujimoto} S.,  et~al., 2019, \mn@doi [\apj] {10.3847/1538-4357/ab480f}, \href
  {https://ui.adsabs.harvard.edu/abs/2019ApJ...887..107F} {887, 107}

\bibitem[\protect\citeauthoryear{{Fujimoto} et~al.,}{{Fujimoto}
  et~al.}{2020}]{Fujimoto20}
{Fujimoto} S.,  et~al., 2020, \mn@doi [\apj] {10.3847/1538-4357/ab94b3}, \href
  {https://ui.adsabs.harvard.edu/abs/2020ApJ...900....1F} {900, 1}

\bibitem[\protect\citeauthoryear{{Gallerani}, {Pallottini}, {Feruglio},
  {Ferrara}, {Maiolino}, {Vallini}, {Riechers}  \& {Pavesi}}{{Gallerani}
  et~al.}{2018}]{Gallerani18}
{Gallerani} S.,  {Pallottini} A.,  {Feruglio} C.,  {Ferrara} A.,  {Maiolino}
  R.,  {Vallini} L.,  {Riechers} D.~A.,   {Pavesi} R.,  2018, \mn@doi [\mnras]
  {10.1093/mnras/stx2458}, \href
  {https://ui.adsabs.harvard.edu/abs/2018MNRAS.473.1909G} {473, 1909}

\bibitem[\protect\citeauthoryear{{Genzel} et~al.,}{{Genzel}
  et~al.}{2015}]{Genzel15}
{Genzel} R.,  et~al., 2015, \mn@doi [\apj] {10.1088/0004-637X/800/1/20}, \href
  {https://ui.adsabs.harvard.edu/abs/2015ApJ...800...20G} {800, 20}

\bibitem[\protect\citeauthoryear{{Ginolfi} et~al.,}{{Ginolfi}
  et~al.}{2017}]{Ginolfi17}
{Ginolfi} M.,  et~al., 2017, \mn@doi [\mnras] {10.1093/mnras/stx712}, \href
  {https://ui.adsabs.harvard.edu/abs/2017MNRAS.468.3468G} {468, 3468}

\bibitem[\protect\citeauthoryear{{Ginolfi} et~al.,}{{Ginolfi}
  et~al.}{2020}]{Ginolfi20}
{Ginolfi} M.,  et~al., 2020, \mn@doi [\aap] {10.1051/0004-6361/202038284},
  \href {https://ui.adsabs.harvard.edu/abs/2020A&A...643A...7G} {643, A7}

\bibitem[\protect\citeauthoryear{{Glover} \& {Clark}}{{Glover} \&
  {Clark}}{2016}]{Glover16}
{Glover} S. C.~O.,  {Clark} P.~C.,  2016, \mn@doi [\mnras]
  {10.1093/mnras/stv2863}, \href
  {https://ui.adsabs.harvard.edu/abs/2016MNRAS.456.3596G} {456, 3596}

\bibitem[\protect\citeauthoryear{{Goulding} et~al.,}{{Goulding}
  et~al.}{2018}]{Goulding18}
{Goulding} A.~D.,  et~al., 2018, \mn@doi [\apj] {10.3847/1538-4357/aab040},
  \href {https://ui.adsabs.harvard.edu/abs/2018ApJ...856....4G} {856, 4}

\bibitem[\protect\citeauthoryear{{Hamann} et~al.,}{{Hamann}
  et~al.}{2017}]{Hamann17}
{Hamann} F.,  et~al., 2017, \mn@doi [\mnras] {10.1093/mnras/stw2387}, \href
  {https://ui.adsabs.harvard.edu/abs/2017MNRAS.464.3431H} {464, 3431}

\bibitem[\protect\citeauthoryear{{Harrison}}{{Harrison}}{2017}]{Harrison17}
{Harrison} C.~M.,  2017, \mn@doi [Nature Astronomy] {10.1038/s41550-017-0165},
  \href {http://adsabs.harvard.edu/abs/2017NatAs...1E.165H} {1, 0165}

\bibitem[\protect\citeauthoryear{{Harrison} et~al.,}{{Harrison}
  et~al.}{2016}]{Harrison16Alm}
{Harrison} C.~M.,  et~al., 2016, \mn@doi [\mnras] {10.1093/mnrasl/slw001},
  \href {http://adsabs.harvard.edu/abs/2016MNRAS.457L.122H} {457, L122}

\bibitem[\protect\citeauthoryear{{Hayward} \& {Hopkins}}{{Hayward} \&
  {Hopkins}}{2017}]{Hayward17}
{Hayward} C.~C.,  {Hopkins} P.~F.,  2017, \mn@doi [\mnras]
  {10.1093/mnras/stw2888}, \href
  {https://ui.adsabs.harvard.edu/abs/2017MNRAS.465.1682H} {465, 1682}

\bibitem[\protect\citeauthoryear{{Herrera-Camus} et~al.,}{{Herrera-Camus}
  et~al.}{2021}]{Herrera-Camus21}
{Herrera-Camus} R.,  et~al., 2021, \mn@doi [\aap]
  {10.1051/0004-6361/202039704}, \href
  {https://ui.adsabs.harvard.edu/abs/2021A&A...649A..31H} {649, A31}

\bibitem[\protect\citeauthoryear{{Hickox}, {Mullaney}, {Alexander}, {Chen},
  {Civano}, {Goulding}  \& {Hainline}}{{Hickox} et~al.}{2014}]{Hickox14}
{Hickox} R.~C.,  {Mullaney} J.~R.,  {Alexander} D.~M.,  {Chen} C.-T.~J.,
  {Civano} F.~M.,  {Goulding} A.~D.,   {Hainline} K.~N.,  2014, \mn@doi [\apj]
  {10.1088/0004-637X/782/1/9}, \href
  {http://adsabs.harvard.edu/abs/2014ApJ...782....9H} {782, 9}

\bibitem[\protect\citeauthoryear{{Hirschmann}, {Dolag}, {Saro}, {Bachmann},
  {Borgani}  \& {Burkert}}{{Hirschmann} et~al.}{2014}]{Hirschmann14}
{Hirschmann} M.,  {Dolag} K.,  {Saro} A.,  {Bachmann} L.,  {Borgani} S.,
  {Burkert} A.,  2014, \mn@doi [\mnras] {10.1093/mnras/stu1023}, \href
  {https://ui.adsabs.harvard.edu/abs/2014MNRAS.442.2304H} {442, 2304}

\bibitem[\protect\citeauthoryear{{Hodge} et~al.,}{{Hodge}
  et~al.}{2016}]{Hodge16}
{Hodge} J.~A.,  et~al., 2016, \mn@doi [\apj] {10.3847/1538-4357/833/1/103},
  \href {http://adsabs.harvard.edu/abs/2016ApJ...833..103H} {833, 103}

\bibitem[\protect\citeauthoryear{{Hopkins}, {Kere{\v{s}}}, {O{\~n}orbe},
  {Faucher-Gigu{\`e}re}, {Quataert}, {Murray}  \& {Bullock}}{{Hopkins}
  et~al.}{2014}]{Hopkins14}
{Hopkins} P.~F.,  {Kere{\v{s}}} D.,  {O{\~n}orbe} J.,  {Faucher-Gigu{\`e}re}
  C.-A.,  {Quataert} E.,  {Murray} N.,   {Bullock} J.~S.,  2014, \mn@doi
  [\mnras] {10.1093/mnras/stu1738}, \href
  {https://ui.adsabs.harvard.edu/abs/2014MNRAS.445..581H} {445, 581}

\bibitem[\protect\citeauthoryear{{Ikarashi} et~al.,}{{Ikarashi}
  et~al.}{2015}]{Ikarashi15}
{Ikarashi} S.,  et~al., 2015, \mn@doi [\apj] {10.1088/0004-637X/810/2/133},
  \href {http://adsabs.harvard.edu/abs/2015ApJ...810..133I} {810, 133}

\bibitem[\protect\citeauthoryear{{Ishikawa}, {Goulding}, {Zakamska}, {Hamann},
  {Vayner}, {Veilleux}  \& {Wylezalek}}{{Ishikawa} et~al.}{2021}]{Ishikawa21}
{Ishikawa} Y.,  {Goulding} A.~D.,  {Zakamska} N.~L.,  {Hamann} F.,  {Vayner}
  A.,  {Veilleux} S.,   {Wylezalek} D.,  2021, \mn@doi [\mnras]
  {10.1093/mnras/stab137}, \href
  {https://ui.adsabs.harvard.edu/abs/2021MNRAS.502.3769I} {502, 3769}

\bibitem[\protect\citeauthoryear{{Jiao}, {Zhao}, {Zhu}, {Lu}, {Gao}  \&
  {Zhang}}{{Jiao} et~al.}{2017}]{Jiao17}
{Jiao} Q.,  {Zhao} Y.,  {Zhu} M.,  {Lu} N.,  {Gao} Y.,   {Zhang} Z.-Y.,  2017,
  \mn@doi [\apjl] {10.3847/2041-8213/aa6f0f}, \href
  {https://ui.adsabs.harvard.edu/abs/2017ApJ...840L..18J} {840, L18}

\bibitem[\protect\citeauthoryear{{Jolly}, {Knudsen}  \& {Stanley}}{{Jolly}
  et~al.}{2020}]{Jolly20}
{Jolly} J.-B.,  {Knudsen} K.~K.,   {Stanley} F.,  2020, \mn@doi [\mnras]
  {10.1093/mnras/staa2908}, \href
  {https://ui.adsabs.harvard.edu/abs/2020MNRAS.499.3992J} {499, 3992}

\bibitem[\protect\citeauthoryear{{Jones}, {Maiolino}, {Caselli}  \&
  {Carniani}}{{Jones} et~al.}{2019}]{Jones19}
{Jones} G.~C.,  {Maiolino} R.,  {Caselli} P.,   {Carniani} S.,  2019, \mn@doi
  [\aap] {10.1051/0004-6361/201936989}, \href
  {https://ui.adsabs.harvard.edu/abs/2019A&A...632L...7J} {632, L7}

\bibitem[\protect\citeauthoryear{{Jones}, {Maiolino}, {Caselli}  \&
  {Carniani}}{{Jones} et~al.}{2020}]{Jones20}
{Jones} G.~C.,  {Maiolino} R.,  {Caselli} P.,   {Carniani} S.,  2020, \mn@doi
  [\mnras] {10.1093/mnras/staa2689}, \href
  {https://ui.adsabs.harvard.edu/abs/2020MNRAS.498.4109J} {498, 4109}

\bibitem[\protect\citeauthoryear{{Jones}, {Maiolino}, {Circosta}, {Scholtz},
  {Carniani}  \& {Fudamoto}}{{Jones} et~al.}{2022}]{Jones22}
{Jones} G.~C.,  {Maiolino} R.,  {Circosta} C.,  {Scholtz} J.,  {Carniani} S.,
  {Fudamoto} Y.,  2022, arXiv e-prints, \href
  {https://ui.adsabs.harvard.edu/abs/2022arXiv221013370J} {p. arXiv:2210.13370}

\bibitem[\protect\citeauthoryear{{Kaasinen} et~al.,}{{Kaasinen}
  et~al.}{2020}]{Kaasinen20}
{Kaasinen} M.,  et~al., 2020, \mn@doi [\apj] {10.3847/1538-4357/aba438}, \href
  {https://ui.adsabs.harvard.edu/abs/2020ApJ...899...37K} {899, 37}

\bibitem[\protect\citeauthoryear{{King} \& {Nixon}}{{King} \&
  {Nixon}}{2015}]{King15}
{King} A.,  {Nixon} C.,  2015, \mn@doi [\mnras] {10.1093/mnrasl/slv098}, \href
  {http://adsabs.harvard.edu/abs/2015MNRAS.453L..46K} {453, L46}

\bibitem[\protect\citeauthoryear{{Klindt}, {Alexander}, {Rosario}, {Lusso}  \&
  {Fotopoulou}}{{Klindt} et~al.}{2019}]{Klindt+19}
{Klindt} L.,  {Alexander} D.~M.,  {Rosario} D.~J.,  {Lusso} E.,   {Fotopoulou}
  S.,  2019, \mn@doi [\mnras] {10.1093/mnras/stz1771}, \href
  {https://ui.adsabs.harvard.edu/abs/2019MNRAS.488.3109K} {488, 3109}

\bibitem[\protect\citeauthoryear{{Kohandel}, {Pallottini}, {Ferrara},
  {Zanella}, {Behrens}, {Carniani}, {Gallerani}  \& {Vallini}}{{Kohandel}
  et~al.}{2019}]{Kohandel19}
{Kohandel} M.,  {Pallottini} A.,  {Ferrara} A.,  {Zanella} A.,  {Behrens} C.,
  {Carniani} S.,  {Gallerani} S.,   {Vallini} L.,  2019, \mn@doi [\mnras]
  {10.1093/mnras/stz1486}, \href
  {https://ui.adsabs.harvard.edu/abs/2019MNRAS.487.3007K} {487, 3007}

\bibitem[\protect\citeauthoryear{{Lambert} et~al.,}{{Lambert}
  et~al.}{2022}]{Lambert22}
{Lambert} T.~S.,  et~al., 2022, arXiv e-prints, \href
  {https://ui.adsabs.harvard.edu/abs/2022arXiv221010023L} {p. arXiv:2210.10023}

\bibitem[\protect\citeauthoryear{{Lamperti} et~al.,}{{Lamperti}
  et~al.}{2021}]{Lamperti21}
{Lamperti} I.,  et~al., 2021, arXiv e-prints, \href
  {https://ui.adsabs.harvard.edu/abs/2021arXiv210902674L} {p. arXiv:2109.02674}

\bibitem[\protect\citeauthoryear{{Lang} et~al.,}{{Lang} et~al.}{2019}]{Lang19}
{Lang} P.,  et~al., 2019, \mn@doi [\apj] {10.3847/1538-4357/ab1f77}, \href
  {https://ui.adsabs.harvard.edu/abs/2019ApJ...879...54L} {879, 54}

\bibitem[\protect\citeauthoryear{{Leroy} et~al.,}{{Leroy}
  et~al.}{2011}]{Leroy11}
{Leroy} A.~K.,  et~al., 2011, \mn@doi [\apj] {10.1088/0004-637X/737/1/12},
  \href {https://ui.adsabs.harvard.edu/abs/2011ApJ...737...12L} {737, 12}

\bibitem[\protect\citeauthoryear{{Li} et~al.,}{{Li} et~al.}{2021a}]{Li+21}
{Li} J.,  et~al., 2021a, \mn@doi [\apjl] {10.3847/2041-8213/ac390d}, \href
  {https://ui.adsabs.harvard.edu/abs/2021ApJ...922L..29L} {922, L29}

\bibitem[\protect\citeauthoryear{Li et~al.,}{Li et~al.}{2021b}]{Li21}
Li J.,  et~al., 2021b, \mn@doi [The Astrophysical Journal Letters]
  {10.3847/2041-8213/ac390d}, 922, L29

\bibitem[\protect\citeauthoryear{{Liang} et~al.,}{{Liang}
  et~al.}{2019}]{Liang19}
{Liang} L.,  et~al., 2019, \mn@doi [\mnras] {10.1093/mnras/stz2134}, \href
  {https://ui.adsabs.harvard.edu/abs/2019MNRAS.489.1397L} {489, 1397}

\bibitem[\protect\citeauthoryear{{Liddle}}{{Liddle}}{2007}]{Liddle07}
{Liddle} A.~R.,  2007, \mn@doi [\mnras] {10.1111/j.1745-3933.2007.00306.x},
  \href {https://ui.adsabs.harvard.edu/abs/2007MNRAS.377L..74L} {377, L74}

\bibitem[\protect\citeauthoryear{{Lindroos}, {Knudsen}, {Vlemmings}, {Conway}
  \& {Mart{\'\i}-Vidal}}{{Lindroos} et~al.}{2015}]{Lindroos2015}
{Lindroos} L.,  {Knudsen} K.~K.,  {Vlemmings} W.,  {Conway} J.,
  {Mart{\'\i}-Vidal} I.,  2015, \mn@doi [\mnras] {10.1093/mnras/stu2344}, \href
  {https://ui.adsabs.harvard.edu/abs/2015MNRAS.446.3502L} {446, 3502}

\bibitem[\protect\citeauthoryear{{Lu} et~al.,}{{Lu} et~al.}{2018}]{Lu18}
{Lu} N.,  et~al., 2018, \mn@doi [\apj] {10.3847/1538-4357/aad3c9}, \href
  {https://ui.adsabs.harvard.edu/abs/2018ApJ...864...38L} {864, 38}

\bibitem[\protect\citeauthoryear{{Lynden-Bell}}{{Lynden-Bell}}{1969}]{LyndenBell69}
{Lynden-Bell} D.,  1969, \mn@doi [\nat] {10.1038/223690a0}, \href
  {http://adsabs.harvard.edu/abs/1969Natur.223..690L} {223, 690}

\bibitem[\protect\citeauthoryear{{Magdis} et~al.,}{{Magdis}
  et~al.}{2011}]{Magdis11}
{Magdis} G.~E.,  et~al., 2011, \mn@doi [\apjl] {10.1088/2041-8205/740/1/L15},
  \href {https://ui.adsabs.harvard.edu/abs/2011ApJ...740L..15M} {740, L15}

\bibitem[\protect\citeauthoryear{{Magnelli} et~al.,}{{Magnelli}
  et~al.}{2012}]{Magnelli12}
{Magnelli} B.,  et~al., 2012, \mn@doi [\aap] {10.1051/0004-6361/201118312},
  \href {https://ui.adsabs.harvard.edu/abs/2012A&A...539A.155M} {539, A155}

\bibitem[\protect\citeauthoryear{{Maiolino} et~al.,}{{Maiolino}
  et~al.}{2012}]{Maiolino12}
{Maiolino} R.,  et~al., 2012, \mn@doi [\mnras]
  {10.1111/j.1745-3933.2012.01303.x}, \href
  {https://ui.adsabs.harvard.edu/abs/2012MNRAS.425L..66M} {425, L66}

\bibitem[\protect\citeauthoryear{{McAlpine}, {Bower}, {Harrison}, {Crain},
  {Schaller}, {Schaye}  \& {Theuns}}{{McAlpine} et~al.}{2017}]{McAlpine17}
{McAlpine} S.,  {Bower} R.~G.,  {Harrison} C.~M.,  {Crain} R.~A.,  {Schaller}
  M.,  {Schaye} J.,   {Theuns} T.,  2017, \mn@doi [\mnras]
  {10.1093/mnras/stx658}, \href
  {http://adsabs.harvard.edu/abs/2017MNRAS.468.3395M} {468, 3395}

\bibitem[\protect\citeauthoryear{{Merloni}, {Rudnick}  \& {Di
  Matteo}}{{Merloni} et~al.}{2004}]{Merloni04}
{Merloni} A.,  {Rudnick} G.,   {Di Matteo} T.,  2004, \mn@doi [\mnras]
  {10.1111/j.1365-2966.2004.08382.x}, \href
  {http://adsabs.harvard.edu/abs/2004MNRAS.354L..37M} {354, L37}

\bibitem[\protect\citeauthoryear{{Nikolic}}{{Nikolic}}{2009}]{Nikolic09}
{Nikolic} B.,  2009, arXiv e-prints, \href
  {https://ui.adsabs.harvard.edu/abs/2009arXiv0912.2317N} {p. arXiv:0912.2317}

\bibitem[\protect\citeauthoryear{{Offner}, {Bisbas}, {Bell}  \&
  {Viti}}{{Offner} et~al.}{2014}]{Offner14}
{Offner} S.~S.~R.,  {Bisbas} T.~G.,  {Bell} T.~A.,   {Viti} S.,  2014, \mn@doi
  [\mnras] {10.1093/mnrasl/slu013}, \href
  {https://ui.adsabs.harvard.edu/abs/2014MNRAS.440L..81O} {440, L81}

\bibitem[\protect\citeauthoryear{{Pallottini}, {Ferrara}, {Gallerani},
  {Vallini}, {Maiolino}  \& {Salvadori}}{{Pallottini}
  et~al.}{2017}]{Pallotini17}
{Pallottini} A.,  {Ferrara} A.,  {Gallerani} S.,  {Vallini} L.,  {Maiolino} R.,
    {Salvadori} S.,  2017, \mn@doi [\mnras] {10.1093/mnras/stw2847}, \href
  {https://ui.adsabs.harvard.edu/abs/2017MNRAS.465.2540P} {465, 2540}

\bibitem[\protect\citeauthoryear{{Papadopoulos} \& {Greve}}{{Papadopoulos} \&
  {Greve}}{2004}]{Papadopoulos04}
{Papadopoulos} P.~P.,  {Greve} T.~R.,  2004, \mn@doi [\apjl] {10.1086/426059},
  \href {https://ui.adsabs.harvard.edu/abs/2004ApJ...615L..29P} {615, L29}

\bibitem[\protect\citeauthoryear{{Papadopoulos}, {Bisbas}  \&
  {Zhang}}{{Papadopoulos} et~al.}{2018}]{Papadopoulos18}
{Papadopoulos} P.~P.,  {Bisbas} T.~G.,   {Zhang} Z.-Y.,  2018, \mn@doi [\mnras]
  {10.1093/mnras/sty1077}, \href
  {https://ui.adsabs.harvard.edu/abs/2018MNRAS.478.1716P} {478, 1716}

\bibitem[\protect\citeauthoryear{{Perrotta}, {Hamann}, {Zakamska},
  {Alexandroff}, {Rupke}  \& {Wylezalek}}{{Perrotta} et~al.}{2019}]{Perrotta19}
{Perrotta} S.,  {Hamann} F.,  {Zakamska} N.~L.,  {Alexandroff} R.~M.,  {Rupke}
  D.,   {Wylezalek} D.,  2019, \mn@doi [\mnras] {10.1093/mnras/stz1993}, \href
  {https://ui.adsabs.harvard.edu/abs/2019MNRAS.488.4126P} {488, 4126}

\bibitem[\protect\citeauthoryear{{Planck Collaboration} et~al.,}{{Planck
  Collaboration} et~al.}{2014}]{Planck13}
{Planck Collaboration} et~al., 2014, \mn@doi [\aap]
  {10.1051/0004-6361/201321591}, \href
  {https://ui.adsabs.harvard.edu/abs/2014A%26A...571A..16P} {571, A16}

\bibitem[\protect\citeauthoryear{{Pointon}, {Kacprzak}, {Nielsen}, {Muzahid},
  {Murphy}, {Churchill}  \& {Charlton}}{{Pointon} et~al.}{2019}]{Pointon19}
{Pointon} S.~K.,  {Kacprzak} G.~G.,  {Nielsen} N.~M.,  {Muzahid} S.,  {Murphy}
  M.~T.,  {Churchill} C.~W.,   {Charlton} J.~C.,  2019, \mn@doi [\apj]
  {10.3847/1538-4357/ab3b0e}, \href
  {https://ui.adsabs.harvard.edu/abs/2019ApJ...883...78P} {883, 78}

\bibitem[\protect\citeauthoryear{{Rees}, {Begelman}, {Blandford}  \&
  {Phinney}}{{Rees} et~al.}{1982}]{Rees82}
{Rees} M.~J.,  {Begelman} M.~C.,  {Blandford} R.~D.,   {Phinney} E.~S.,  1982,
  \mn@doi [\nat] {10.1038/295017a0}, \href
  {https://ui.adsabs.harvard.edu/abs/1982Natur.295...17R} {295, 17}

\bibitem[\protect\citeauthoryear{{R{\'e}my-Ruyer} et~al.,}{{R{\'e}my-Ruyer}
  et~al.}{2014}]{Remy-Ruyer14}
{R{\'e}my-Ruyer} A.,  et~al., 2014, \mn@doi [\aap]
  {10.1051/0004-6361/201322803}, \href
  {https://ui.adsabs.harvard.edu/abs/2014A&A...563A..31R} {563, A31}

\bibitem[\protect\citeauthoryear{{Reuter} et~al.,}{{Reuter}
  et~al.}{2020}]{Reuter20}
{Reuter} C.,  et~al., 2020, \mn@doi [\apj] {10.3847/1538-4357/abb599}, \href
  {https://ui.adsabs.harvard.edu/abs/2020ApJ...902...78R} {902, 78}

\bibitem[\protect\citeauthoryear{{Rohlfs} \& {Wilson}}{{Rohlfs} \&
  {Wilson}}{1996}]{Rohlfs96}
{Rohlfs} K.,  {Wilson} T.~L.,  1996, {Tools of Radio Astronomy}

\bibitem[\protect\citeauthoryear{{Ross} et~al.,}{{Ross} et~al.}{2015}]{Ross15}
{Ross} N.~P.,  et~al., 2015, \mn@doi [\mnras] {10.1093/mnras/stv1710}, \href
  {https://ui.adsabs.harvard.edu/abs/2015MNRAS.453.3932R} {453, 3932}

\bibitem[\protect\citeauthoryear{{Sanderson}, {Prescott}, {Christensen},
  {Fynbo}  \& {M{\o}ller}}{{Sanderson} et~al.}{2021}]{sand21}
{Sanderson} K.~N.,  {Prescott} M. M.~K.,  {Christensen} L.,  {Fynbo} J.,
  {M{\o}ller} P.,  2021, \mn@doi [\apj] {10.3847/1538-4357/ac3077}, \href
  {https://ui.adsabs.harvard.edu/abs/2021ApJ...923..252S} {923, 252}

\bibitem[\protect\citeauthoryear{{Sandstrom} et~al.,}{{Sandstrom}
  et~al.}{2013}]{Sandstrom13}
{Sandstrom} K.~M.,  et~al., 2013, \mn@doi [\apj] {10.1088/0004-637X/777/1/5},
  \href {https://ui.adsabs.harvard.edu/abs/2013ApJ...777....5S} {777, 5}

\bibitem[\protect\citeauthoryear{{Schawinski}, {Koss}, {Berney}  \&
  {Sartori}}{{Schawinski} et~al.}{2015}]{Schawinski15}
{Schawinski} K.,  {Koss} M.,  {Berney} S.,   {Sartori} L.~F.,  2015, \mn@doi
  [\mnras] {10.1093/mnras/stv1136}, \href
  {http://adsabs.harvard.edu/abs/2015MNRAS.451.2517S} {451, 2517}

\bibitem[\protect\citeauthoryear{{Scholtz} et~al.,}{{Scholtz}
  et~al.}{2018}]{Scholtz18}
{Scholtz} J.,  et~al., 2018, \mn@doi [\mnras] {10.1093/mnras/stx3177}, \href
  {http://adsabs.harvard.edu/abs/2018MNRAS.475.1288S} {475, 1288}

\bibitem[\protect\citeauthoryear{{Scholtz} et~al.,}{{Scholtz}
  et~al.}{2020}]{Scholtz20}
{Scholtz} J.,  et~al., 2020, \mn@doi [\mnras] {10.1093/mnras/staa030}, \href
  {https://ui.adsabs.harvard.edu/abs/2020MNRAS.492.3194S} {492, 3194}

\bibitem[\protect\citeauthoryear{{Scholtz} et~al.,}{{Scholtz}
  et~al.}{2021}]{Scholtz21}
{Scholtz} J.,  et~al., 2021, \mn@doi [\mnras] {10.1093/mnras/stab1631}, \href
  {https://ui.adsabs.harvard.edu/abs/2021MNRAS.505.5469S} {505, 5469}

\bibitem[\protect\citeauthoryear{{Schreiber} et~al.,}{{Schreiber}
  et~al.}{2015}]{Schreiber15}
{Schreiber} C.,  et~al., 2015, \mn@doi [\aap] {10.1051/0004-6361/201425017},
  \href {http://adsabs.harvard.edu/abs/2015A%26A...575A..74S} {575, A74}

\bibitem[\protect\citeauthoryear{{Schreiber}, {Elbaz}, {Pannella}, {Ciesla},
  {Wang}  \& {Franco}}{{Schreiber} et~al.}{2018}]{Schreiber18}
{Schreiber} C.,  {Elbaz} D.,  {Pannella} M.,  {Ciesla} L.,  {Wang} T.,
  {Franco} M.,  2018, \mn@doi [\aap] {10.1051/0004-6361/201731506}, \href
  {https://ui.adsabs.harvard.edu/abs/2018A&A...609A..30S} {609, A30}

\bibitem[\protect\citeauthoryear{Schwarz}{Schwarz}{1978}]{Schwarz78}
Schwarz G.,  1978, Ann. Statist., 6, 461

\bibitem[\protect\citeauthoryear{{Scoville} et~al.,}{{Scoville}
  et~al.}{2014}]{Scoville14}
{Scoville} N.,  et~al., 2014, \mn@doi [\apj] {10.1088/0004-637X/783/2/84},
  \href {https://ui.adsabs.harvard.edu/abs/2014ApJ...783...84S} {783, 84}

\bibitem[\protect\citeauthoryear{{Segers}, {Schaye}, {Bower}, {Crain},
  {Schaller}  \& {Theuns}}{{Segers} et~al.}{2016}]{Segers16}
{Segers} M.~C.,  {Schaye} J.,  {Bower} R.~G.,  {Crain} R.~A.,  {Schaller} M.,
  {Theuns} T.,  2016, \mn@doi [\mnras] {10.1093/mnrasl/slw111}, \href
  {https://ui.adsabs.harvard.edu/abs/2016MNRAS.461L.102S} {461, L102}

\bibitem[\protect\citeauthoryear{{Silk} \& {Rees}}{{Silk} \&
  {Rees}}{1998}]{Silk98}
{Silk} J.,  {Rees} M.~J.,  1998, \aap, \href
  {https://ui.adsabs.harvard.edu/abs/1998A&A...331L...1S} {331, L1}

\bibitem[\protect\citeauthoryear{{Simpson} et~al.,}{{Simpson}
  et~al.}{2015}]{Simpson15}
{Simpson} J.~M.,  et~al., 2015, \mn@doi [\apj] {10.1088/0004-637X/807/2/128},
  \href {http://adsabs.harvard.edu/abs/2015ApJ...807..128S} {807, 128}

\bibitem[\protect\citeauthoryear{{Soltan}}{{Soltan}}{1982}]{Soltan82}
{Soltan} A.,  1982, \mn@doi [\mnras] {10.1093/mnras/200.1.115}, \href
  {https://ui.adsabs.harvard.edu/abs/1982MNRAS.200..115S} {200, 115}

\bibitem[\protect\citeauthoryear{{Somerville} \& {Dav{\'e}}}{{Somerville} \&
  {Dav{\'e}}}{2015}]{Somerville15}
{Somerville} R.~S.,  {Dav{\'e}} R.,  2015, \mn@doi [\araa]
  {10.1146/annurev-astro-082812-140951}, \href
  {https://ui.adsabs.harvard.edu/abs/2015ARA&A..53...51S} {53, 51}

\bibitem[\protect\citeauthoryear{{Spilker} et~al.,}{{Spilker}
  et~al.}{2016}]{Spilker16}
{Spilker} J.~S.,  et~al., 2016, \mn@doi [\apj] {10.3847/0004-637X/826/2/112},
  \href {http://adsabs.harvard.edu/abs/2016ApJ...826..112S} {826, 112}

\bibitem[\protect\citeauthoryear{{Spilker} et~al.,}{{Spilker}
  et~al.}{2018}]{Spilker18}
{Spilker} J.~S.,  et~al., 2018, \mn@doi [Science] {10.1126/science.aap8900},
  \href {https://ui.adsabs.harvard.edu/abs/2018Sci...361.1016S} {361, 1016}

\bibitem[\protect\citeauthoryear{{Stanley}, {Jolly}, {K{\"o}nig}  \&
  {Knudsen}}{{Stanley} et~al.}{2019}]{Stanley19}
{Stanley} F.,  {Jolly} J.~B.,  {K{\"o}nig} S.,   {Knudsen} K.~K.,  2019,
  \mn@doi [\aap] {10.1051/0004-6361/201834530}, \href
  {https://ui.adsabs.harvard.edu/abs/2019A&A...631A..78S} {631, A78}

\bibitem[\protect\citeauthoryear{{Tadaki} et~al.,}{{Tadaki}
  et~al.}{2017}]{Tadaki17}
{Tadaki} K.-i.,  et~al., 2017, \mn@doi [\apj] {10.3847/1538-4357/834/2/135},
  \href {http://adsabs.harvard.edu/abs/2017ApJ...834..135T} {834, 135}

\bibitem[\protect\citeauthoryear{Tazzari}{Tazzari}{2017}]{uvplot_tazzari}
Tazzari M.,  2017, mtazzari/uvplot, \mn@doi{10.5281/zenodo.1003113}, \url
  {https://doi.org/10.5281/zenodo.1003113}

\bibitem[\protect\citeauthoryear{{Tielens} \& {Hollenbach}}{{Tielens} \&
  {Hollenbach}}{1985}]{Tielens85}
{Tielens} A.~G.~G.~M.,  {Hollenbach} D.,  1985, \mn@doi [\apj]
  {10.1086/163111}, \href
  {https://ui.adsabs.harvard.edu/abs/1985ApJ...291..722T} {291, 722}

\bibitem[\protect\citeauthoryear{{Travascio} et~al.,}{{Travascio}
  et~al.}{2020}]{Travascio20}
{Travascio} A.,  et~al., 2020, \mn@doi [\aap] {10.1051/0004-6361/201936197},
  \href {https://ui.adsabs.harvard.edu/abs/2020A&A...635A.157T} {635, A157}

\bibitem[\protect\citeauthoryear{{Umehata} et~al.,}{{Umehata}
  et~al.}{2019}]{Umehata19}
{Umehata} H.,  et~al., 2019, \mn@doi [Science] {10.1126/science.aaw5949}, \href
  {https://ui.adsabs.harvard.edu/abs/2019Sci...366...97U} {366, 97}

\bibitem[\protect\citeauthoryear{{Valentino} et~al.,}{{Valentino}
  et~al.}{2018}]{Valentino18}
{Valentino} F.,  et~al., 2018, \mn@doi [\apj] {10.3847/1538-4357/aaeb88}, \href
  {https://ui.adsabs.harvard.edu/abs/2018ApJ...869...27V} {869, 27}

\bibitem[\protect\citeauthoryear{{Vayner}, {Zakamska}, {Wright}, {Armus},
  {Murray}  \& {Walth}}{{Vayner} et~al.}{2021}]{Vayner21}
{Vayner} A.,  {Zakamska} N.,  {Wright} S.~A.,  {Armus} L.,  {Murray} N.,
  {Walth} G.,  2021, \mn@doi [\apj] {10.3847/1538-4357/ac2b9e}, \href
  {https://ui.adsabs.harvard.edu/abs/2021ApJ...923...59V} {923, 59}

\bibitem[\protect\citeauthoryear{{Vogelsberger} et~al.,}{{Vogelsberger}
  et~al.}{2014}]{Vogelsberger14}
{Vogelsberger} M.,  et~al., 2014, \mn@doi [\mnras] {10.1093/mnras/stu1536},
  \href {http://adsabs.harvard.edu/abs/2014MNRAS.444.1518V} {444, 1518}

\bibitem[\protect\citeauthoryear{{Walter}, {Wei{\ss}}, {Downes}, {Decarli}  \&
  {Henkel}}{{Walter} et~al.}{2011}]{Walter11}
{Walter} F.,  {Wei{\ss}} A.,  {Downes} D.,  {Decarli} R.,   {Henkel} C.,  2011,
  \mn@doi [\apj] {10.1088/0004-637X/730/1/18}, \href
  {https://ui.adsabs.harvard.edu/abs/2011ApJ...730...18W} {730, 18}

\bibitem[\protect\citeauthoryear{{Wardlow} et~al.,}{{Wardlow}
  et~al.}{2018}]{Wardlow18}
{Wardlow} J.~L.,  et~al., 2018, \mn@doi [\mnras] {10.1093/mnras/sty1526}, \href
  {https://ui.adsabs.harvard.edu/abs/2018MNRAS.479.3879W} {479, 3879}

\bibitem[\protect\citeauthoryear{{Wright} et~al.,}{{Wright}
  et~al.}{2010}]{Wright10}
{Wright} E.~L.,  et~al., 2010, \mn@doi [\aj] {10.1088/0004-6256/140/6/1868},
  \href {https://ui.adsabs.harvard.edu/abs/2010AJ....140.1868W} {140, 1868}

\bibitem[\protect\citeauthoryear{Zhao et~al.,}{Zhao et~al.}{2020}]{Zhao20}
Zhao Y.,  et~al., 2020, \mn@doi [The Astrophysical Journal]
  {10.3847/1538-4357/ab75eb}, 892, 145

\makeatother
\end{thebibliography}

%%%%%%%%%%%%%%%%%%%%%%%%%%%%%%%%%%%%%%%%%%%%%%%%%%

%%%%%%%%%%%%%%%%%%%%%%%%%%%%%%%%%%%%%%%%%%%%%%%%%%

%%%%%%%%%%%%%%%%% APPENDICES %%%%%%%%%%%%%%%%%%%%%

\appendix
\section{Full stacking results}

\begin{landscape}
\begin{figure}
    \includegraphics[width=0.99\paperwidth]{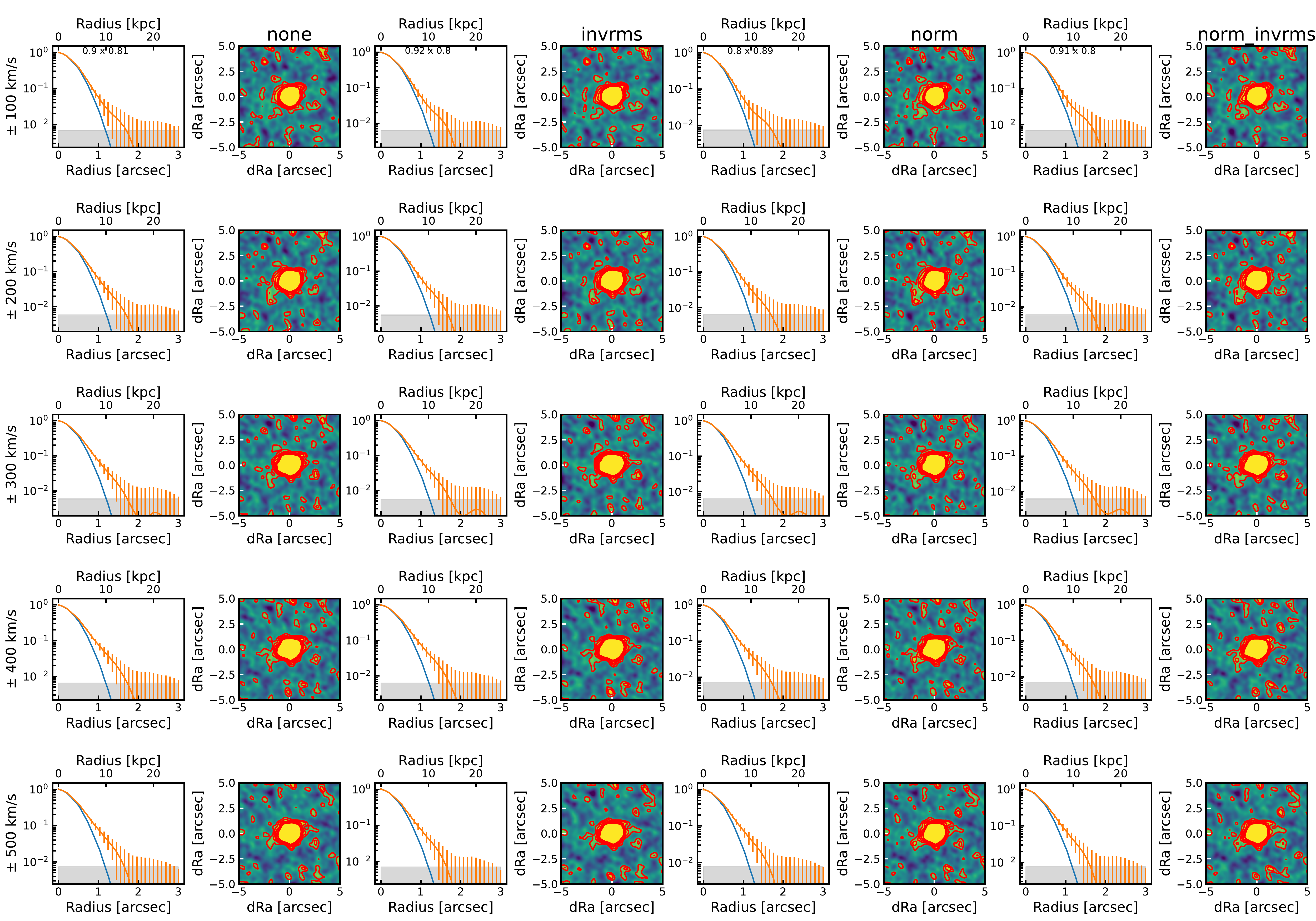}
   \caption{Results of the CO(7-6) stacking and associated brightness profiles. Moment-0 and radial brightness profiles of the stacked cubes. Each column corresponds to different stacking normalisation: none, inverse rms, normalised to peak value and a combination of previous both normalisations. Each row corresponds to moment-0 map from different velocity range. Moment-0 maps:  The collapsed emission is shown with contours of $\sigma$ contours, starting at $1\sigma$. Brightness profiles: The orange line shows normalised brightness profiles with 1 $\sigma$ errors. The blue line shows the effective restoring beam of the stacked cube. 
    }
   \label{fig:Stacking_CO}
\end{figure} 
\end{landscape}

\begin{landscape}
\begin{figure}
    \includegraphics[width=0.99\paperwidth]{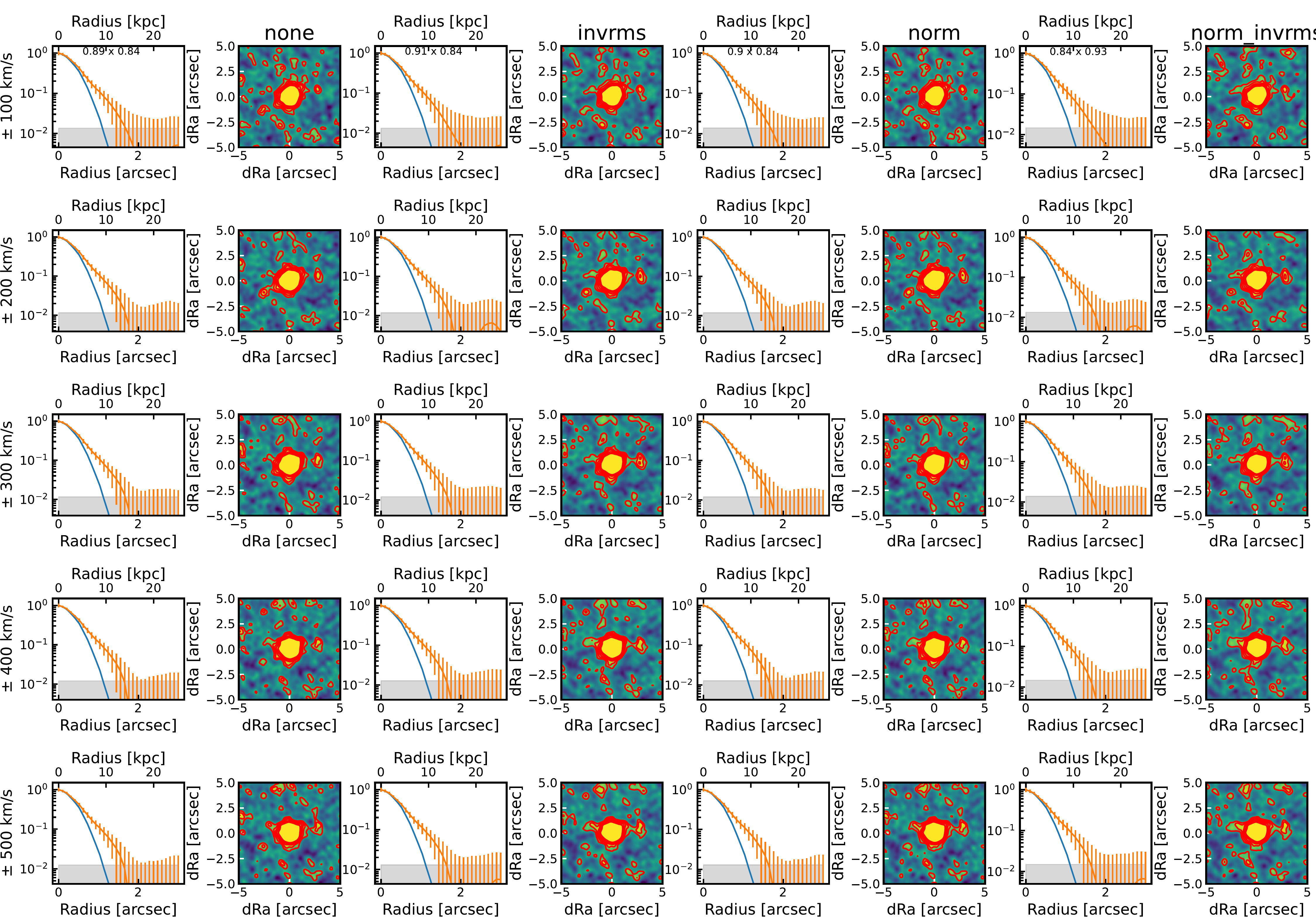}
   \caption{ Results of the [\ion{C}{i}](2-1) stacking and associated brightness profiles. Moment-0 and radial brightness profiles of the stacked cubes. Each column corresponds to different stacking normalisation: none, inverse rms, normalised to peak value and a combination of previous both normalisations. Each row corresponds to moment-0 map from a different velocity range. Moment-0 maps:  The collapsed emission is shown with contours of $\pm \sigma$ contours, starting at $\pm 1\sigma$ (positive and negative $\sigma$ are in dashed and solid contours, respectively). Brightness profiles: The orange line shows normalised brightness profiles with 1 $\sigma$ errors. The blue line shows the effective restoring beam of the stacked cube. 
    }
   \label{fig:Stacking_[CI]}
\end{figure} 
\end{landscape}

\begin{landscape}
\begin{figure}
    \includegraphics[width=0.99\paperwidth]{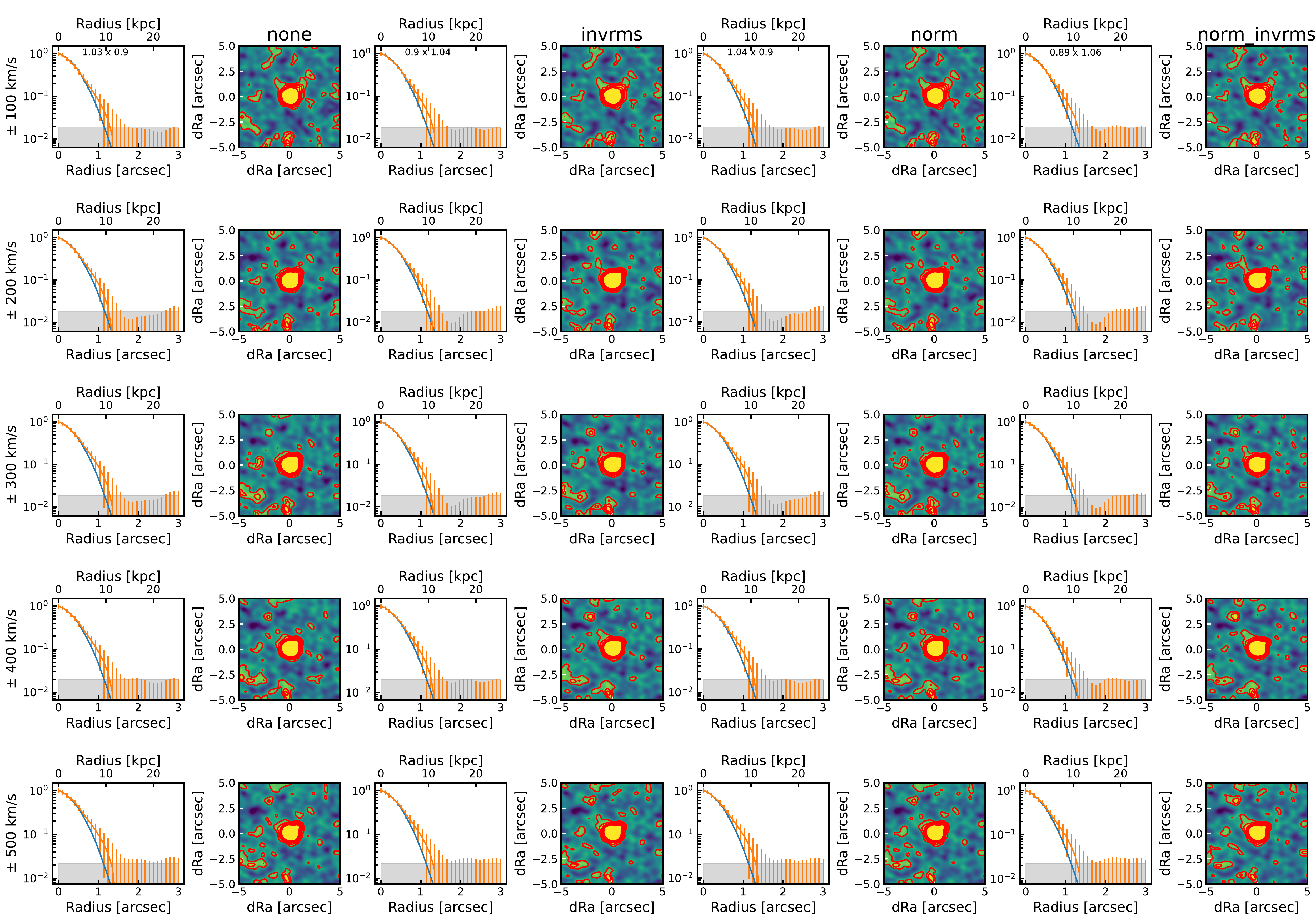}
   \caption{ Results of the H$_2$O stacking and associated brightness profiles. Moment-0 and radial brightness profiles of the stacked cubes. Each column corresponds to different stacking normalisation: none, inverse rms, normalised to peak value and a combination of previous both normalisations. Each row corresponds to moment-0 map from different velocity range. Moment-0 maps:  The collapsed emission is shown with contours of $\pm \sigma$ contours, starting at $\pm 1\sigma$ (positive and negative $\sigma$ are in dashed and solid contours, respectively). Brightness profiles: The orange line shows normalised brightness profiles with 1 $\sigma$ errors. The blue line shows the effective restoring beam of the stacked cube. 
    }
   \label{fig:Stacking_H2O]}
\end{figure} 
\end{landscape}

\begin{figure*}
    \includegraphics[width=0.8\paperwidth]{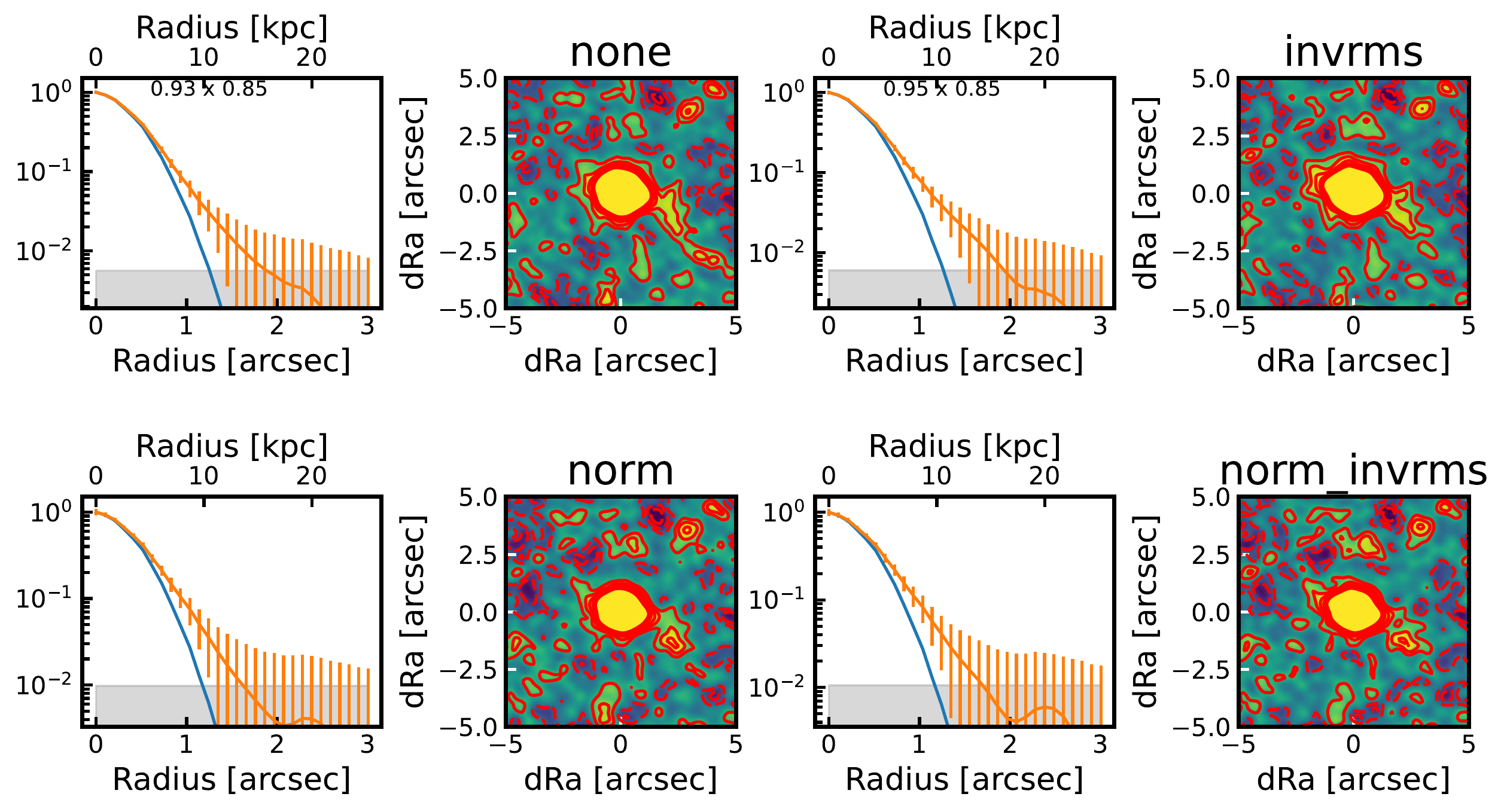}
   \caption{ Results of the Continum stacking and associated brightness profiles. The stacked image and radial brightness profiles of the stacked cubes. Each column corresponds to different stacking normalisation: none, inverse rms, normalised to peak value and a combination of previous both normalisations. Each row corresponds to moment-0 map from different velocity range. Dust images:  Dust emission map is shown with contours of $\pm \sigma$ contours, starting at $\pm 1\sigma$ (positive and negative $\sigma$ are in dashed and solid contours, respectively). Brightness profiles: The orange line shows normalised brightness profiles with 1 $\sigma$ errors. The blue line shows the effective restoring beam of the stacked cube. 
    }
   \label{fig:Stacking_cont]}
\end{figure*}

In Figures \ref{fig:Stacking_CO}, \ref{fig:Stacking_[CI]}, \ref{fig:Stacking_H2O]} \& \ref{fig:Stacking_cont]} we present the the full results of the image-plane stacking across all weighting scheme and five velocity ranges. 

%%%%%%%%%%%%%%%%%%%%%%%%%%%%%%%%%%%%%%%%%%%%%%%%%%

% Don't change these lines
\bsp	% typesetting comment
\label{lastpage}
\end{document}